&pdflatex
\documentclass[a4paper,authoryear,3p]{elsarticle}

\usepackage{amssymb}
\usepackage{amsmath}
\usepackage{nicefrac}
\usepackage{tabu}

\journal{Elsevier}

\begin{document}

\let\originalleft\left
\let\originalright\right
\def\left#1{\mathopen{}\originalleft#1}
\def\right#1{\originalright#1\mathclose{}}

\begin{frontmatter}

\title{High-order accurate finite-volume formulations for the pressure gradient force in layered ocean models}

\author[mit,nasa]{Darren Engwirda\corref{cor1}}
\ead{engwirda@mit.edu}
\cortext[cor1]{Corresponding author\@. Tel.: +1-212-678-5521}

\author[nasa]{Maxwell Kelley}
\ead{maxwell.kelley@nasa.gov}
\author[mit]{John Marshall}
\ead{jmarsh@mit.edu}

\address[mit]{Department of Earth, Atmospheric and Planetary Sciences, Massachusetts Institute of Technology, 54-918, 77 Massachusetts Avenue, Cambridge, MA 02139-4307, USA}

\address[nasa]{NASA Goddard Institute for Space Studies, 2880 Broadway, New York, NY 10025 USA}

\begin{abstract} 
The development of a set of high-order accurate finite-volume formulations for evaluation of the pressure gradient force in layered ocean models is described. A pair of new schemes are presented, both based on an integration of the \textit{contact} pressure force about the perimeter of an associated momentum control-volume. The two proposed methods differ in their choice of control-volume geometries. High-order accurate numerical integration techniques are employed in both schemes to account for non-linearities in the underlying equation-of-state definitions and thermodynamic profiles, and details of an associated vertical interpolation and quadrature scheme are discussed in detail. Numerical experiments are used to confirm the consistency of the two formulations, and it is demonstrated that the new methods maintain hydrostatic and thermobaric equilibrium in the presence of strongly-sloping layer-wise geometry, non-linear equation-of-state definitions and non-uniform vertical stratification profiles. Additionally, one scheme is shown to maintain high levels of consistency in the presence of non-linear thermodynamic stratification. Use of the new pressure gradient force formulations for hybrid vertical coordinate and/or terrain-following general circulation models is discussed.
\end{abstract}


\end{frontmatter}

\allowdisplaybreaks

\section{Introduction}
\label{section_introduction}

The development of flexible layered ocean models, capable of adapting to the complex vertical structure associated with stratified geophysical flows, represents an important ongoing numerical challenge in global climate modelling and numerical weather prediction. Compared to conventional \textit{fixed-grid} formulations, layered models, in which the fluid is subdivided into a set of curvilinear layers, offer an opportunity to improve the fidelity with which vertical ocean transport processes are represented \citep{griffies2000spurious}. In this study, the issue of constructing a consistent and accurate numerical formulation for evaluation of the horizontal pressure gradient force in arbitrarily layered ocean models is discussed in detail. While seemingly innocuous, the development of stable and consistent pressure gradient formulations presents significant numerical challenges, due to the complex interplay between non-linearities in the underlying fluid equation-of-state, the depth-wise stratification profiles, and the sloping geometry of the discrete fluid layers themselves. 

The paper is organised as follows: in Section~\ref{section_model_description}, a simplified framework for layered ocean model development is presented, with the equations-of-motion expressed in terms of an arbitrary vertical coordinate. The genesis of the numerical instabilities associated with conventional formulations for the horizontal pressure gradient force are described, and a review of several existing techniques presented. In Sections~\ref{section_layerwise_method} and \ref{section_rectilinear_method}, new finite-volume type formulations are described, focusing in detail on the construction of flexible, high-order accurate numerical integration procedures compatible with generalised non-linear equations-of-state and stratification profiles. Experimental results obtained using these new schemes are presented in Section~\ref{section_layerwise_results}, along with a comparison of the relative performance of the two formulations for several \textit{ocean-at-rest} type benchmark problems.

\section{A simplified layered ocean model}

\label{section_model_description}

Following \citet{bleck2002oceanic}, \citet{adcroft2006methods}, \citet{higdon2002one,higdon2005two} and \citet{leclair2011coordinate}, the layered, hydrostatic and non-Boussinesq equations of motion for a rotating geophysical fluid can be expressed in terms of a generalised vertical coordinate $s$ as a set of five prognostic conservation laws -- two for the horizontal velocity components, two for a pair of thermodynamic variables, an evolution equation for a layerwise \textit{thickness} variable, and a diagnostic expression for the equation-of-state of the fluid. In this study, a fully Lagrangian-type representation is employed, requiring that the flow rate $\dot{s}$ normal to surfaces of constant $s$ be identically zero. Such a constraint implies dynamic motion of the coordinate surfaces themselves, with the thickness of the fluid layers evolving in time due to vertical motion.  
\begin{gather}
\label{eqn_momentum_1}
\partial_{t}\left(\mathbf{u}_{h}\right) + (\mathbf{u}_{h}\cdot\nabla_{s})\mathbf{u}_{h} + f\mathbf{u}_{h}^{\bot} = \nabla_{s}\left(\Phi\right) + \rho^{-1} \nabla_{s}\left(p\right) + \mathbf{F}_{\mathbf{u}_{h}}\,, \text{\qquad} \partial_{s}(\Phi) = \rho^{-1}
\\[2ex]
\label{eqn_layer_1}
\partial_{t}(\delta) + \nabla_{s}\cdot\left(\delta \mathbf{u}_{h}\right) = F_{\delta}
\\[2ex]
\label{eqn_temp_1}
\partial_{t}\left(\delta T\right) + \nabla_{s}\cdot\left(\delta \mathbf{u}_{h} T\right) = F_{T}
\\[2ex]
\label{eqn_salt_1}
\partial_{t}\left(\delta S\right) + \nabla_{s}\cdot\left(\delta \mathbf{u}_{h} S\right) = F_{S}
\end{gather}
Here $\mathbf{u}_{h} = (u,v)$ is the horizontal velocity field, $\mathbf{u}_{h}^{\bot} = (-v,u)$, $f$ is the Coriolis parameter, $\Phi=gz$ is the geopotential, where $g$ is the acceleration due to gravity and $z$ is the height from a reference surface, $\delta$ is the \textit{pressure-thickness} associated with a given layer of fluid, and $T$ and $S$ are the scalar temperature and salinity distributions, respectively. Note that the specific choice of thermodynamic pairing is dependent on the equation of state used, with, for example, potential temperature and practical salinity $\left(T,S\right)=\left(\theta,S_{p}\right)$ used in a number of existing thermodynamic models \citep{wright1997equation}, while recent formulations \citep{mcdougall2011getting} necessitate a switch to the conservative temperature and absolute salinity pair $\left(T,S\right)=\left(\Theta,S_{\text{A}}\right)$. The forcing terms $\mathbf{F}_{\mathbf{u}_{h}}$, $F_{p}$, $F_{T}$ and $F_{S}$ incorporate any additional sources and sinks of horizontal force, freshwater, and thermodynamic response, in addition to the effect of generalised diffusion/mixing on both the momentum and thermodynamic variables, respectively. The fluid density $\rho = f\left(T,S,p\right)$ is diagnosed via a general non-linear equation of state, and the geopotential $\Phi=gz$ is expressed in terms of hydrostatic balance. Note that the differential operator $\partial_{t}$ denotes a derivative with respect to time, $\partial_{s}$ denotes a derivative with respect to the generalised vertical coordinate $s$, and $\nabla_{s} = \left(\partial_{x},\partial_{y},0\right)$ is the layerwise gradient operator, taken along surfaces of constant $s$. Expressions for the transport of passive tracers an be added to this system via the inclusion of additional advection-diffusion equations of the form of (Eqn.~\ref{eqn_salt_1}). 

\subsection{Existing formulations for the horizontal pressure gradient operator}

\medskip

Numerical issues related to the discretisation of the horizontal pressure gradient force have long plagued the development of layered ocean models. These numerical errors typically manifest as spurious horizontal accelerations -- causing the model to erroneously `drift' away from the desired equilibrium state over time. The genesis of such difficulties can be explained by examining the interaction of the two differential operators associated with the pressure gradient force in Eqn.~\ref{eqn_momentum_1}:
\begin{gather}
\label{eqn_pgf}
\operatorname{PGF} = \nabla_{s}\left(\Phi\right) + \rho^{-1} \nabla_{s}\left(p\right)
\end{gather}
Given particular (conventional) choices of vertical coordinate, namely $s=z$ or $s=p$, the form of the pressure gradient operator can be simplified, with one of the two gradient terms ($\nabla_{s}\left(\Phi\right)$ and $\rho^{-1} \nabla_{s}\left(p\right)$) evaluating to zero. Specifically, in conventional height-based coordinates $\nabla_{s}\left(\Phi\right) = \nabla_{z}\left(\Phi\right) = 0$, while in a pressure-based coordinate system $\rho^{-1} \nabla_{s}\left(p\right) = \rho^{-1} \nabla_{p}\left(p\right) = 0$. Unfortunately, this exact cancellation is not preserved when adopting arbitrary vertical coordinate systems appropriate for layered ocean modelling, such as terrain-following coordinates and/or time- and space-dependent Lagrangian representations. In such cases, a straight-forward discretisation of the two gradient operators in (Eqn.~\ref{eqn_pgf}) can lead to inconsistencies, with the interaction of the numerical truncation errors associated with each gradient term leading to inexact cancellation. Noting that the magnitude of these two terms is typically large compared to the dynamical signal \citep{adcroft2008finite}, it can be understood that residual errors in the evaluation of the pressure gradient force can lead to non-negligible spurious horizontal motion. This behaviour is exacerbated when the fluid layers are steeply sloping and the imposed thermodynamic stratification profiles are highly non-uniform.

Conventionally, layered isopycnic-type models \citep{bleck2002oceanic} have sought to exploit the so-called \textit{Mont\-gomery-potential} form of the horizontal pressure gradient operator. Setting $M = \nicefrac{p}{\rho} + \Phi$, the horizontal acceleration can be transformed as follows:
\begin{gather}
\label{eqn_pgf_2}
\operatorname{PGF} = \nabla_{s}\left(M\right) + p \nabla_{s}\left(\rho^{-1}\right)
\end{gather}
Note that in an exact density-following coordinate system ($s=\rho$), the second term in Eqn.~\ref{eqn_pgf_2} can be seen to vanish, with $p \nabla_{s}\left(\rho^{-1}\right) = p \nabla_{\rho}\left(\rho^{-1}\right) = 0$. While such a result is attractive from a theoretical standpoint, it should be noted that practical isopycnic-type models do not typically adopt a coordinate system based on the exact in-situ densities, preferring instead hybrid potential-density-based representations, with height-based transitions employed near layer outcropping \citep{bleck2002oceanic}. Nonetheless, it can be argued that use of the Montgomery potential form serves to mitigate associated numerical errors, through a minimisation of the magnitude of the second gradient term $p \nabla_{s}\left(\rho^{-1}\right)$. In practice, such considerations are known not to be fully satisfactory, with studies of models based on layerwise finite-difference type discretisations of the Montgomery potential reported to suffer from serious issues of instability \citep{adcroft2008finite}.  

Alternatively, finite-volume type discretisations for the pressure gradient operator have also been proposed, seeking to properly account for the interaction between layerwise geometry, pressure-compressibility and thermodynamic stratification effects through the evaluation of a suitable set of boundary integrals. Specifically, the net horizontal pressure force acting on a layerwise control-volume can be computed by integration of the so-called \textit{contact} pressure force acting at the boundary of each control-volume. Such approaches have been pursued fruitfully by a number of authors, including \citet{lin1997finite}, \citet{shchepetkin2005regional} and \citet{adcroft2008finite}. More recently, such finite-volume approaches have been supplemented by so-called semi-analytic methods, providing improved accuracy and efficiency. In \citet{adcroft2008finite}, a hydrostatically consistent integration method is presented, where \textit{exact} vertical profiles of geopotential are computed using analytic integration rules. It was shown that, under certain simplifying assumptions, use of the semi-analytic pressure gradient formulation led to significant improvements in the stability and consistency of an associated layered isopycnic-type model. 
 
\section{The semi-analytic finite-volume formulation}

\medskip

\begin{figure}[t]

\caption{Sloping quadrilateral control-volumes $\Omega_{i+\nicefrac{1}{2},k}$ associated with the layer-wise pressure gradient formulation. The staggered control-volumes $\Omega_{i+\nicefrac{1}{2},k}$ are formed by joining the top and bottom edge-midpoints of adjacent mass grid-cells. Note that $\Omega_{i+\nicefrac{1}{2},k}$ achieves a piece-wise linear approximation to bottom topography.}

\begin{center}
\includegraphics[width=.275\textwidth]{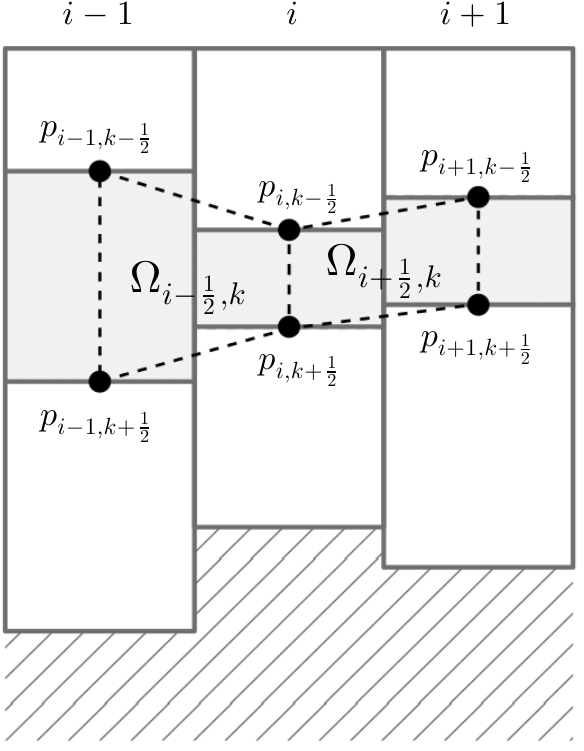} \qquad\qquad
\includegraphics[width=.275\textwidth]{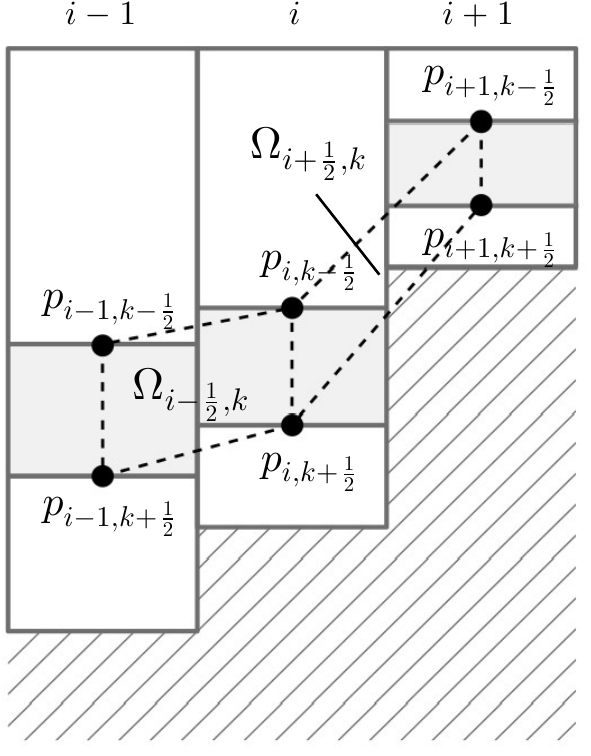}
\end{center}

\label{figure_cv_omega}

\end{figure}

A finite-volume scheme for evaluation of the horizontal pressure gradient terms in the momentum equation (Eqn.~\ref{eqn_momentum_1}) can be formulated through a summation of the \textit{contact} pressure force acting at the boundaries of the piecewise linear control-volumes associated with discrete momentum components. In integral form:
\begin{gather}
\label{eqn_pressure_integral_0}
\operatorname{PGF} = 
\frac{1}{\Delta x_{i+\frac{1}{2},k}}
\frac{1}{\Delta p_{i+\frac{1}{2},k}} 
\oint_{\partial\Omega}\Phi\,\mathrm{d}p
\\[2ex]
\label{eqn_pressure_integral_1}
\oint_{\partial\Omega}\Phi\,\mathrm{d}p = 
\int_{p_{b_{r}}}^{p_{t_{r}}}\Phi\,\mathrm{d}p + 
\int_{p_{t_{r}}}^{p_{t_{l}}}\Phi\,\mathrm{d}p+
\int_{p_{t_{l}}}^{p_{b_{l}}}\Phi\,\mathrm{d}p +
\int_{p_{b_{l}}}^{p_{b_{r}}}\Phi\,\mathrm{d}p
\end{gather}
where the contour integral has been split into the four segments -- taken in a counter-clockwise order -- associated with the edges of the two-dimensional quadrilateral control-volume $\Omega$ associated with a given horizontal velocity variable, as illustrated in Figure~\ref{figure_cv_omega}. Note that a \textit{staggered} horizontal arrangement is employed, with the velocity control-volumes $\Omega$ located between the midpoints of associated \textit{mass} layers. This arrangement is consistent with a conventional C-type grid-staggering, in which velocity variables are offset from adjacent thermodynamic and layer-thickness quantities. Adopting the conventional nomenclature, each horizontal velocity component $u_{i+\nicefrac{1}{2},k}$ is staggered between the thermodynamic variables $(T_{i,k},S_{i,k})$ and $(T_{i+1,k},S_{i+1,k})$ and layer pressure-thickness quantities $\delta_{i,k}$ and $\delta_{i+1,k}$, where the $(i,k)$ indices denote the horizontal and vertical directions, respectively. The fluid pressure is staggered in the vertical direction, and is stored at the layer interfaces $p_{i,k+\nicefrac{1}{2}}$ for each column. Note that the depth-wise index $k$ increases downward from the fluid surface. In this study, the \textit{mass} grid-cells are referred to as the primary \textit{fluid-columns}, while the velocity grid-points are termed the staggered or \textit{dual} control-volumes.

Noting that the geopotential $\Phi$ is a non-linear function of both the fluid pressure $p$ and thermodynamic variables $T(p)$, $S(p)$, it is clear that discretisation of the contact pressure expressions (Eqn.~\ref{eqn_pressure_integral_0}) represents a significant numerical challenge. Specifically, it is required that any numerical scheme designed to discretise the line integral terms (Eqn.~\ref{eqn_pressure_integral_1}) faithfully account for this complex set of non-linear dependencies. Failure to adequately capture such effects can cause an imbalance in the contact pressure force computed along each segment of the control-volume boundary -- leading to the well-known issues of horizontal pressure gradient force error and instability, as outlined in the previous section.

In \citet{adcroft2008finite}, the so-called \textit{semi-analytic} formulation was proposed, where, under the assumption of a prescribed equation-of-state and piecewise constant thermodynamic profiles, an analytic solution to the hydrostatic relationship was derived. Specifically, given a simplified equation-of-state of the form \citep{wright1997equation}:
\begin{gather}
\label{eqn_wright_eos}
\rho^{-1}\left(T,S,p\right) = A\left(T,S\right) + \frac{\lambda\left(T,S\right)}{P\left(T,S\right) + p}
\end{gather}  
where $A\left(T,S\right) = A$, $\lambda\left(T,S\right) = \lambda$ and $P\left(T,S\right) = P$ (under the assumption of piecewise constant $T,S$ profiles), as per \citet{adcroft2008finite}, the exact variation in geopotential can be computed analytically as:
\begin{gather}
\label{eqn_adcroft_hydro}
\begin{split}
\Phi(p_{t}) - \Phi(p_{b}) = \int_{p_{t}}^{p_{b}} \rho^{-1}\left(T,S,p\right) \,\mathrm{d}p &= \Delta p \left( A + \frac{\lambda}{2\epsilon(P + \bar{p})}\ln \left|\frac{1+\epsilon}{1-\epsilon}\right| \right) 
\\[2ex]
&= \Delta p \left( A + \frac{\lambda}{(P + \bar{p})} \, \sum_{i=1}^{\infty} \, \frac{\epsilon^{2i-2}}{2i-1} \right)
\end{split}
\end{gather} 
where the expressions have been simplified following the nomenclature of \citet{adcroft2008finite}, such that:
\begin{gather}
\Delta p = p_{b} - p_{t}\,, \text{\qquad} \bar{p} = \frac{p_{b}+p_{t}}{2}\,, \text{\qquad} \epsilon = \frac{\Delta p}{2(P+\bar{p})}
\end{gather}
Using Eqn.~\ref{eqn_adcroft_hydro}, expressions for the pressure gradient force itself can be derived. The contributions from the left- and right-hand edges of the control-volume can be computed exactly, through an additional analytic integration of Eqn.~\ref{eqn_adcroft_hydro} over the respective edge segments:
\begin{gather}
\label{eqn_adcroft_integral}
\begin{split}
\int_{p_{b}}^{p_{t}} \Phi \,\mathrm{d}p &= \Delta p \left( \Phi_{b} + \frac{1}{2}A\Delta p + \lambda \left( 1 - \frac{1-\epsilon}{2\epsilon} \ln \left|\frac{1+\epsilon}{1-\epsilon}\right| \right) \right) 
\\[2ex] 
&= \Delta p \left( \Phi_{b} + \frac{1}{2}A\Delta p + \lambda \left( \epsilon - (\epsilon^{2}-\epsilon^{3}) \, \sum_{i=1}^{\infty} \, \frac{\epsilon^{2i-2}}{2i+1} \right) \right)
\end{split}
\end{gather}
As per \citet{adcroft2008finite}, evaluation of the infinite series in Eqn.~\ref{eqn_adcroft_hydro} and \ref{eqn_adcroft_integral} can be computed approximately by summing over a finite number of terms. Specifically, evaluation of the first six terms in each series has been reported to lead to approximations accurate to within numerical rounding errors. 

Evaluation of the line integral terms along the sloping upper and lower control-volume edges is significantly less straightforward, due to the horizontal variation in both the fluid pressure-thickness and thermodynamic variables over the cell width. In \citet{adcroft2008finite}, it is remarked that these terms cannot readily be evaluated analytically, and a numerical integration approach is pursued instead. Specifically, Eqn.~\ref{eqn_adcroft_hydro} is used to compute the \textit{exact} increment in geopotential height over the layer-thickness, with a horizontal interpolation of the coefficients in the equation-of-state (Eqn.~\ref{eqn_wright_eos}) employed to account for variations in the thermodynamic quantities along the layer. Given a distribution of geopotential values over the control-volume edges, the resulting contact pressure force integrals can be computed as weighted sums, as per standard numerical quadrature techniques \citep{golub1969calculation,abramowitz1964handbook}.

A detailed comparison of the performance of the semi-analytic finite-volume formulation and a conventional Montgomery-potential approach was presented in \citet{adcroft2008finite}. The semi-analytic formulation was shown to outperform the conventional approach, offering significant improvements to both the consistency and accuracy of results obtained using a layered isopycnic ocean model \citep{hallberg1996buoyancy,hallberg2005thermobaric}. Specifically, the semi-analytic scheme was shown to exactly preserve hydrostatic consistency in simplified ocean conditions, and to suppress grid-scale oscillations generated using the potential-based approach.

\section{Method I: A layer-wise finite-volume formulation}

\label{section_layerwise_method}

While offering significant improvements over conventional two-term pressure gradient formulations, the flexibility of the original semi-analytic finite-volume approach of \citet{adcroft2008finite} is limited by its underlying assumptions. Specifically, the \textit{exact} analytic integration results developed in \citet{adcroft2008finite} rely on a number of factors, including: (i) the assumption of piecewise constant thermodynamic profiles over the layer thicknesses, and (ii) the use of a simplified equation-of-state \citep{wright1997equation}. These constraints encourage the development of more generalised methods.

Dispensing with analytic integration, a new, flexible finite-volume formulation for evaluation of the pressure gradient force based solely on high-order \textit{numerical} integration techniques is proposed. Such an approach is designed to extend the semi-analytic formulation presented previously to support more realistic, non-uniform thermodynamic profiles, and arbitrary equation-of-state definitions. 

\subsection{Preliminaries: Numerical integration}
\label{section_numerical_integration}

\medskip

In contrast to \citet{adcroft2008finite}, a numerical evaluation of the line-integral terms in Eqn.~\ref{eqn_pressure_integral_1} is sought. This integration is a two-step process, firstly seeking to assemble the column-wise profiles of geopotential $\Phi_{i}$, through integration of the hydrostatic relationship, before evaluating the \textit{contact} pressure integrals given in Eqn.~\ref{eqn_pressure_integral_1}. A summation of the contact pressures about the four sides of each control-volume $\Omega_{i+\frac{1}{2},k}$ leads to an approximation of the pressure gradient force, as per Eqn.~\ref{eqn_pressure_integral_0}. The vertical profile of geopotential $\Phi_{i}$ in each fluid column $i$ is given by:
\begin{gather}
\label{eqn_hydro_exact}
\Phi_{i}(p) - \Phi_{b} = \int_{p_{b}}^{p} \rho^{-1}\,\mathrm{d}p
\end{gather}
where $\Phi_{b}$ and $p_{b}$ are the values of geopotential and fluid pressure at the base of the column, respectfully. The fluid density $\rho$ is assumed to be a fully non-linear equation-of-state, such that $\rho=f(T_{i},S_{i},p)$, where $T_{i}(p)$ and $S_{i}(p)$ are arbitrary vertical profiles of the thermodynamic quantities within the associated column.

Recalling that $\Phi_{i}(p)$ varies non-linearly over the stack of control-volumes $\Omega_{i+\nicefrac{1}{2},k}$ in each column, a suitably accurate numerical integration of Eqn.~\ref{eqn_hydro_exact} is sought. Such behaviour can be realised using an appropriate set of \textit{numerical-quadrature} rules \citep{golub1969calculation,abramowitz1964handbook} of sufficiently high-order. The use of quadrature rules requires the integrand in Eqn.~\ref{eqn_hydro_exact} -- the reciprocal of the fluid density -- be evaluated at a discrete set of \textit{quadrature-points} distributed over the layer thicknesses. Recalling that evaluation of the contact pressure integrals requires a two-step integration process, a non-standard form of numerical quadrature is employed in this study, designed to allow the same set of function evaluations to be re-used within each pass of the nested integration steps. Noting that the density of seawater is typically specified in terms of complex non-linear functions \citep{mcdougall2011getting}, a minimisation of function evaluations is an important consideration when seeking to construct efficient numerical schemes.

The geopotential profile $\Phi_{i,k}(p)$ in the $k$-th layer of the $i$-th column can be found by integrating Eqn.~\ref{eqn_hydro_exact}, where a suitable polynomial approximation to $\rho^{-1}$ is exploited:
\begin{gather}
\label{eqn_hydro_integral}
\begin{split}
\Phi_{i,k}(p)-\Phi_{i,k+\frac{1}{2}} = \int_{p_{i,k+\frac{1}{2}}}^{p} \rho^{-1}\,\mathrm{d}p 
&\simeq \Delta p\int_{0}^{\xi} a_{1} + a_{2}\xi + \dots + a_{n}\xi^{n-1}\,\mathrm{d}\xi
\\[1ex]
&\simeq \Delta p\,\left(a_{1}\xi + \frac{1}{2}a_{2}\xi^{2} + \dots + \frac{1}{n}a_{n}\xi^{n}\right)
\end{split}
\end{gather}
Here the vertical variation in $\Phi_{i,k}$ is computed for a given layer spanning between the upper and lower pressure levels $p_{i,k-\nicefrac{1}{2}}$ and $p_{i,k+\nicefrac{1}{2}}$, such that the layer thickness $\Delta p = p_{i,k+\nicefrac{1}{2}}-p_{i,k-\nicefrac{1}{2}}$. Additionally, the non-dimensional variable $\xi$ has been introduced to map the integration region onto the uniform segment $\xi \in [0,1]$. Such a mapping can be expressed through the following transformation:
\begin{gather}
\label{eqn_local_mapping}
p_{i,k}(\xi) = p_{i,k-\frac{1}{2}}+\xi\left(p_{i,k+\frac{1}{2}}-p_{i,k-\frac{1}{2}}\right)
\end{gather}
Integration of the hydrostatic expression is completed by determining the polynomial coefficients $a_{1},a_{2},\dots,a_{n}\in\mathbb{R}$ in Eqn.~\ref{eqn_hydro_integral}. This process is based on the construction of a polynomial approximation to $\rho^{-1}(\xi)$ on $\xi\in[0,1]$, and requires the sampling of the fluid density $\rho(T_{i,k}(\xi_{l}),S_{i,k}(\xi_{l}),p_{i,k}(\xi_{l}))$ at a sequence of \textit{integration-points} $\xi_{l}\in [0,1]$ distributed over the integration segment. This curve-fitting procedure is described in detail in \ref{appendix_int}. The resulting polynomial coefficients can be expressed as the solution to a set of linear equations:
\begin{gather}
\label{eqn_poly_coeff}
\left[
\begin{matrix}
a_{1}\\[4pt]
a_{2}\\[4pt]
\vdots\\[4pt]
a_{n}\\[4pt]
\end{matrix}
\right] 
= \mathbf{R}^{-1}\,
\left[
\begin{matrix}
\rho^{-1}\Big(T(\xi_{1}),S(\xi_{1}),p(\xi_{1})\Big)\\[4pt]
\rho^{-1}\Big(T(\xi_{2}),S(\xi_{2}),p(\xi_{2})\Big)\\[4pt]
\vdots\\[4pt]
\rho^{-1}\Big(T(\xi_{n}),S(\xi_{n}),p(\xi_{n})\Big)\\[4pt]
\end{matrix}
\right] 
\end{gather}
where $\mathbf{R}^{-1}$ is an $n\times n$ matrix of reconstruction coefficients that are pre-computed for a given quadrature rule. Clearly, the degree of the interpolating polynomial is related to the number of integration points used, with higher-order interpolants corresponding to additional sampling points. In this study we adopt the conventional terminology, referring to a scheme involving $n$ integration points as an $n$-point quadrature rule. 

Note that in Eqn.~\ref{eqn_poly_coeff}, the sampling of the fluid density $\rho(\xi_{l})$ requires a corresponding evaluation of the associated thermodynamic variables $T_{i,k}(\xi_{l})$ and $S_{i,k}(\xi_{l})$. In this study, such values are obtained using high-order piecewise polynomial reconstructions \citep{colella1984piecewise,white2008high,engwirda2016weno} in which a set of vertical polynomial profiles $T_{i,k}(\xi)$ and $S_{i,k}(\xi)$ are \textit{reconstruced} from the associated layer-wise degrees-of-freedom. Specifically, the piecewise-linear (PLM), piecewise-parabolic (PPM) and piecewise-quartic methods (PQM) are considered in the current work, providing a family of high-order accurate, essentially monotonic polynomial reconstructions for the thermodynamical quantities. The fluid pressure $p_{i,k}(\xi)$ is assumed to vary linearly within each control-volume and is obtained at the integration points $\xi_{l}$ through a corresponding bi-linear interpolation scheme.

\subsection{Evaluation of integral terms over the left- \& right-hand segments}

\medskip

Returning to the evaluation of the integral expressions for the pressure gradient force acting over the control-volume $\Omega_{i+\nicefrac{1}{2},k}$, contributions from the left- and right-hand \textit{side} integral terms are first considered:
\begin{gather}
\oint_{\partial\Omega}\Phi\,\mathrm{d}p = 
\underbrace{
\int_{p_{b_{r}}}^{p_{t_{r}}}\Phi\,\mathrm{d}p + 
\int_{p_{t_{l}}}^{p_{b_{l}}}\Phi\,\mathrm{d}p}_{\text{`side' terms}} +
\int_{p_{t_{r}}}^{p_{t_{l}}}\Phi\,\mathrm{d}p +
\int_{p_{b_{l}}}^{p_{b_{r}}}\Phi\,\mathrm{d}p
\end{gather}
Using Eqn.~\ref{eqn_hydro_integral}, the integrated pressure force acting on the left- and right-hand edges of the control volume can be calculated. The variation of geopotential along the left-hand edge of a control-volume in the $k$-th fluid layer can be expressed as:
\begin{gather}
\label{eqn_geo_layer}
\Phi_{i,k}(\xi) = \Delta p\,\left(a_{1}\xi + \frac{1}{2}a_{2}\xi^{2} + \dots + \frac{1}{n}a_{n}\xi^{n}\right) + \Phi_{i,k+\frac{1}{2}}
\end{gather}
where $\Phi_{i,k+\nicefrac{1}{2}}$ is the value of the geopotential at the base of the layer. In the bottom-most layer this value is simply the bottom boundary condition. As per Section~\ref{section_numerical_integration}, the corresponding polynomial coefficients $a_{l}$ can be computed by sampling the equation-of-state $\rho^{-1}(T_{i,k}(\xi_{l}),S_{i,k}(\xi_{l}),p_{i,k}(\xi_{l}))$ at the integration points distributed over the control-volume thickness. Noting that these left- and right-hand integrals are coincident with the centres of the $i$-th and $i+1$-th fluid columns, the associated thermodynamic variables can be computed in a strictly per-column basis -- there is no need for horizontal interpolation. As discussed previously, these values are obtained by evaluating a set of local piecewise polynomial reconstructions $T_{i,k}(\xi),\,S_{i,k}(\xi)$, obtained via a local PLM, PPM or PQM interpolant, at the integration points $\xi_{l}$. The fluid pressure is computed at the integration points via linear interpolation.

Given the variation in $\Phi_{i,k}(p)$ along each edge, the contribution to the pressure gradient force can be computed by performing a second integration for the associated contact pressure:
\begin{gather}
\label{eqn_side_integral}
\int_{p_{i,k-\frac{1}{2}}}^{p_{i,k+\frac{1}{2}}}\Phi\,\mathrm{d}p = (\Delta p)^{2}\,\left(\frac{1}{2}a_{1} + \frac{1}{6}a_{2} + \dots + \frac{1}{n(n+1)}a_{n}\right) + \Delta p\,\Phi_{i,k+\frac{1}{2}}
\end{gather}
where the integration has been evaluated over the full layer thickness $\xi\in [0,1]$. An evaluation of the pressure gradient force contribution on the right-hand side of the control-volume $\Omega_{i+\nicefrac{1}{2},k}$ can be obtained by repeating this procedure for the edge aligned with the $(i+1)$-th column.

\begin{figure}[t]

\caption{Details of the pressure gradient force calculation for the layer-wise scheme. The distribution of geopotential within the staggered control-volume $\Omega_{i+\nicefrac{1}{2},k}$ is computed using numerical integration techniques. The final contact pressure acting on $\Omega_{i+\nicefrac{1}{2},k}$ is calculated by a subsequent integration of the geopotential distribution along the four edge-segments of the control-volume.}

\label{fig_pgf1}

\begin{center}
\includegraphics[width=.70\textwidth]{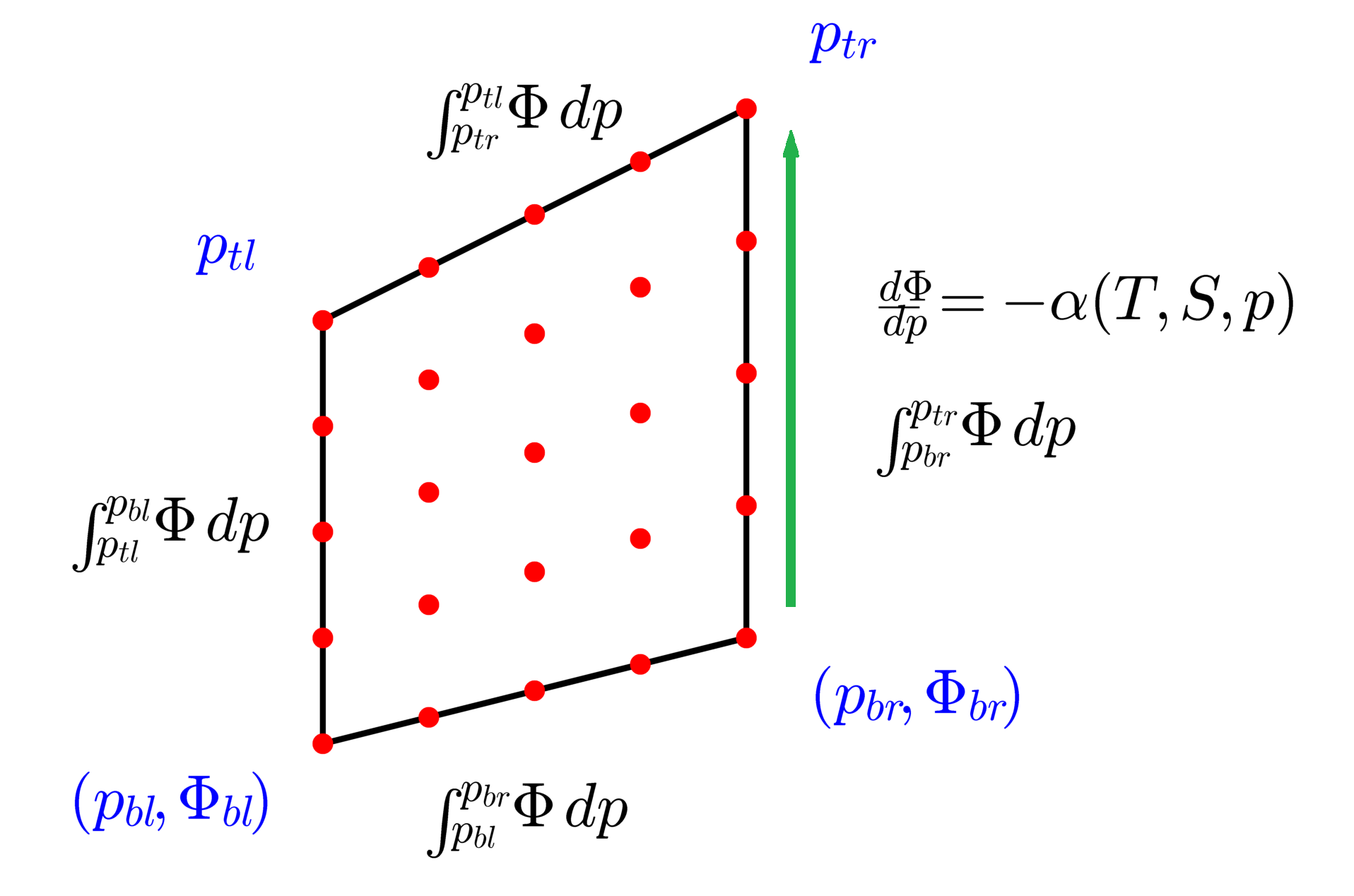}
\end{center}

\end{figure}

\subsection{Evaluation of integral terms on the upper \& lower segments}

\medskip

The contributions to the pressure gradient force from the upper and lower edges of the control-volume $\Omega_{i+\nicefrac{1}{2},k}$ can be computed by integrating the varying geopotential height along the sloping layer interfaces:
\begin{gather}
\oint_{\partial\Omega}\Phi\,\mathrm{d}p = 
\int_{p_{b_{r}}}^{p_{t_{r}}}\Phi\,\mathrm{d}p + 
\int_{p_{t_{l}}}^{p_{b_{l}}}\Phi\,\mathrm{d}p +
\underbrace{
\int_{p_{t_{r}}}^{p_{t_{l}}}\Phi\,\mathrm{d}p +
\int_{p_{b_{l}}}^{p_{b_{r}}}\Phi\,\mathrm{d}p}_{\text{`layer' terms}}
\end{gather}
Contrary to the evaluation of the side integral terms, these calculations are somewhat more involved. Firstly, recalling the arguments presented in Section~\ref{section_numerical_integration}, the pressure gradient force contributions can be computed by numerical quadrature -- sampling the geopotential height at a series of integration points distributed along the upper and lower control-volume edges:
\begin{gather}
\label{eqn_layer_lobatto}
\int_{p_{i+0,k+\frac{1}{2}}}^{p_{i+1,k+\frac{1}{2}}}\Phi\,\mathrm{d}p \simeq \Delta p\,\sum_{l=1}^{n}w_{l}\Phi(x_{l},p_{l})
\end{gather}
Here $\Delta p$ is the horizontal pressure difference along the control-volume edge, the $w_{l}$'s are a set of linear weights associated with a particular choice of quadrature rule, and the points $(x_{l},\,p_{l})$ are the set of integration points distributed along the sloping control-volume edge. Noting that values for the geopotential are already available at the left- and right-hand edges of $\Omega_{i+\nicefrac{1}{2},k}$, due to calculations already performed for the side integral terms, a Lobatto-type quadrature rule \citep{abramowitz1964handbook} is employed in this study, reducing the number of intermediate integration points required to be computed.

The values of geopotential height are calculated at the integration points distributed over the \textit{interior} of the control-volume through additional hydrostatic integration. Specifically, a variant of Eqn.~\ref{eqn_geo_layer} is used to evaluate the geopotential profiles in the control-volume interior, through integration of the hydrostatic relation:
\begin{gather}
\label{eqn_geo_interior}
\Phi_{l,k}(\xi) = \Delta p(x_{l})\,\left(a_{1}(x_{l})\xi + \frac{1}{2}a_{2}(x_{l})\xi^{2} + \dots + \frac{1}{n}a_{n}(x_{l})\xi^{n}\right) + \Phi_{l,k-\frac{1}{2}}
\end{gather}
Note that Eqn.'s~\ref{eqn_geo_layer} and~\ref{eqn_geo_interior} are equivalent, except that, in the latter, an explicit horizontal dependence for both the layer thickness $\Delta p(x_{l})$ and polynomial coefficients $a_{i}(x_{l})$ is accounted for. Evaluation of the pressure thickness $\Delta p(x_{l})$ is unambiguous, with the pressure exactly represented by a bilinear distribution within each control volume $\Omega_{i+\nicefrac{1}{2},k}$. Evaluation of the polynomial coefficients associated with the thermodynamic quantities, though, incorporates an additional level of approximation, with a corresponding horizontal interpolation of thermodynamic quantities required. In this study, these internal values $T(x_{l},p_{l}),\,S(x_{l},p_{l})$ are obtained via a linear interpolation of the associated column-wise reconstructions from the edges of $\Omega_{i+\nicefrac{1}{2},k}$. Note that such a scheme supports high-orders of accuracy in the vertical, but is limited to linear representations in the horizontal. More specifically, this `horizontal' interpolation is actually carried out in a `layer-wise' orientation, and departs from the true horizontal direction when the slope of the layer is non-negligible. This issue will be discussed in detail in subsequent sections. The construction of higher-order accurate horizontal interpolation schemes is a possible avenue for future work.

A somewhat subtle issue relates to the direction of vertical integration for the intermediate profiles $\Phi_{l,k}(p)$. It is tempting to consider an approach in which all geopotential profiles are integrated together, including those for the control-volume edges, starting from the base of the column and working upwards toward the fluid surface. The difficulty with this approach hinges on the formulation of the bottom boundary condition for the intermediate values $\Phi_{l,n_{z}}(p=p_{b})$. Considering the non-linear character of the hydrostatic relationship (Eqn.~\ref{eqn_momentum_1}), it should be noted that it is not consistent for both the bottom pressure and geopotential boundary conditions to vary linearly over the lowest control-volume edge. In fact the correct relationship can only be determined through a consistent integration of the hydrostatic relationship downwards from the fluid surface. 

As such, an alternative multi-stage procedure is employed in this study, in which the set of column-wise geopotential profiles $\Phi_{i,k}(p)$ are first obtained, integrating from the base of each column upwards to the fluid surface. Secondly, a consistent, horizontal geopotential distribution is constructed at the fluid surface for the intermediate profiles $\Phi_{l,1}(p=p_{s})$, by linear interpolation from the adjacent column surface heights. Finally, the intermediate profiles $\Phi_{l,k}(p)$ are computed by integration from the fluid surface downwards towards the bottom boundary. Such a formulation ensures that intermediate profiles of geopotential are computed in a hydrostatically consistent fashion for all horizontal integration points. This two-stage integration process is illustrated in Figure~\ref{fig_pgf1}.

\subsection{Summary of layer-wise pressure gradient formulation}

\medskip

The procedure to evaluate the horizontal pressure gradient force using the layer-wise finite-volume formulation can be summarised in the following steps:
\begin{enumerate}\itemsep=+4pt
\item Compute the set of piecewise polynomial reconstructions in the vertical direction for the thermodynamic variables. Specifically, a set of piecewise polynomial interpolants $T_{i,k}(p),\,S_{i,k}(p)$ are computed for each column in the model.

\item Integrate for the column-centred geopotential values $\Phi_{i,k}(p)$ and compute the pressure force contributions for the control-volume sides. Integration proceeds layer-by-layer from the base of each column, with Eqn.~\ref{eqn_geo_layer} used to obtain values for the geopotential at the layer interfaces. Eqn.~\ref{eqn_side_integral} is used to compute the associated contributions to the pressure gradient force.

\item Construct a surface boundary condition for the intermediate geopotential profiles $\Phi_{k,1}(p=p_{s})$ for all control-volumes. In this study, such values are obtained by linear interpolation from the column surface heights.

\item Integrate for the intermediate geopotential values $\Phi_{l,k}(p)$ and compute the pressure force contributions for the control-volume upper and lower edges. Integration proceeds layer-by-layer from the top of each column, with Eqn.~\ref{eqn_geo_interior} used to obtain values for the geopotential at the interior integration points on layer interfaces. A linear horizontal interpolation for the thermodynamic quantities is performed in this step. Eqn.~\ref{eqn_layer_lobatto} is used to compute the associated contributions to the pressure gradient force.
\end{enumerate}

\section{Method II: A rectilinear finite-volume formulation}

\label{section_rectilinear_method}

While the layer-wise pressure gradient formulation presented previously achieves high-order accuracy in the vertical direction, it is limited by the low-order `horizontal' interpolation scheme used to evaluate terms on the sloping upper and lower control-volume edges. As will be shown in subsequent sections, this effect can lead to issues when the imposed stratification profiles are non-linear and the fluid layers themselves are steeply sloping. As such, an alternative formulation is considered. This second scheme is based on the observation that hydrostatic consistency is easiest to maintain when computations are restricted to \textit{non-staggered} points in the horizontal direction. Specifically, when all hydrostatic integration is carried out at the centre of mass columns, there is no need to perform horizontal interpolation operations, with centred layer-thickness and thermodynamic variables immediately available. The \textit{rectilinear} finite-volume scheme presented in this section seeks to achieve such a discretisation through the selection of an appropriate staggered control-volume geometry.

\subsection{An overlapping axis-aligned control-volume}

\medskip

\begin{figure}[t]

\caption{Overlapping axis-aligned control-volumes $\Gamma_{i+\nicefrac{1}{2},k}$ associated with the rectilinear pressure gradient formulation. The control-volumes $\Gamma_{i+\nicefrac{1}{2},k}$ are axis-aligned rectangles, formed by taking the mean of the layer interface positions of adjacent mass grid-cells. Note that the control-volumes induced by such a strategy may necessarily overlap adjacent fluid layers. In the vicinity of the bottom and surface boundaries, the geometry of the control-volumes $\Gamma_{i+\nicefrac{1}{2},k}$ may be modified to ensure they lie within the fluid interior.}

\label{figure_cv_gamma}

\begin{center}
\includegraphics[width=.275\textwidth]{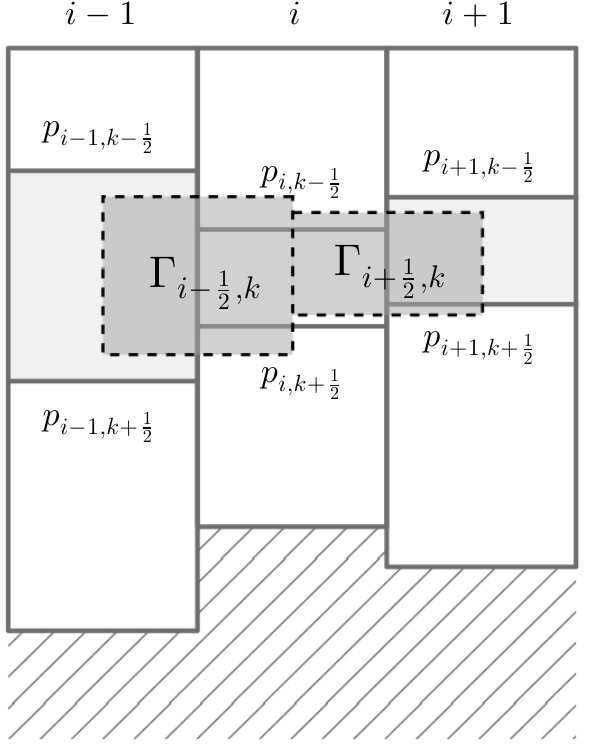} \qquad\qquad
\includegraphics[width=.275\textwidth]{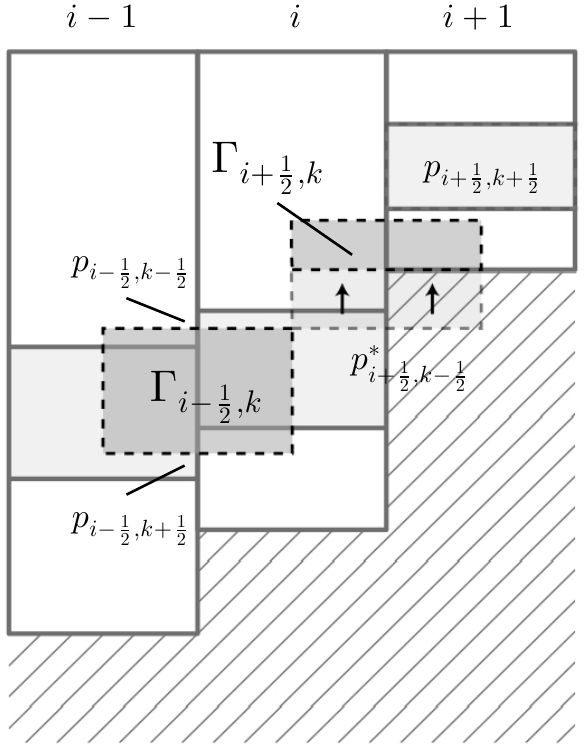}
\end{center}
\end{figure}

In contrast to the sloping quadrilateral control-volumes used in both the semi-analytic formulation of \citet{adcroft2008finite} and the layer-wise methodology presented in Section~\ref{section_layerwise_method}, an alternative axis-aligned control-volume configuration $\Gamma_{i+\nicefrac{1}{2},k}$ is proposed here. Such a geometry is designed to be free of sloping upper and/or lower edge segments, and, as a result, requires an evaluation of the contact pressure acting on the left- and right-hand edges only. As such, an approximation to the pressure gradient term acting over the rectilinear control-volume $\Gamma_{i+\nicefrac{1}{2},k}$ leads to the following integral expressions: 
\begin{gather}
\label{eqn_pressure_integral_2}
\operatorname{PGF} = 
\frac{1}{\Delta x_{i+\frac{1}{2},k}}
\frac{1}{\Delta p_{i+\frac{1}{2},k}} 
\oint_{\partial\Gamma} \Phi\,\mathrm{d}p
\\[2ex]
\label{eqn_pressure_integral_3}
\oint_{\partial\Gamma} \Phi\,\mathrm{d}p = 
\int_{p_{b}}^{p_{t}}\Phi\,\mathrm{d}p +
\int_{p_{t}}^{p_{b}}\Phi\,\mathrm{d}p
\end{gather}
where the contour integral has been split into the two non-trivial segments -- taken in a counter-clockwise order -- corresponding to the left- and right-hand edges of the two-dimensional rectangular control-volume $\Gamma_{i+\nicefrac{1}{2},k}$ associated with a horizontal velocity variable, as illustrated in Figure~\ref{figure_cv_gamma}. Consistent with the layer-wise formulation presented in Section~\ref{section_layerwise_method}, the control-volume $\Gamma_{i+\nicefrac{1}{2},k}$ employs a C-type grid-staggering, with $\Gamma_{i+\nicefrac{1}{2},k}$ sandwiched between a set of thermodynamic and layer-thickness quantities associated with the $i$-th and $(i+1)$-th columns.

In contrast to previous approaches, the control-volume $\Gamma_{i+\nicefrac{1}{2},k}$ is not constrained to lie within a single layer of fluid in the vertical direction, but instead intersects with an overlapping set of layers in the adjacent $i$-th and $(i+1)$-th columns, depending on the particular configuration of relative layer-thicknesses. In some sense, this overlapping finite-volume scheme is related to the class of \textit{truely-horizontal} pressure gradient formulations recently employed in the atmospheric modelling community \citep{zangl2012extending}, where a consistent horizontal pressure gradient is computed by interpolating quantities onto a common height and taking finite-differences. The present scheme can be thought of as a generalised integral form of such approaches, where the pressure gradient force is approximated as the \textit{truely-horizontal} difference between integrated contact pressures acting over a finite control-volume. 

The vertical extent of the control-volume $\Gamma_{i+\nicefrac{1}{2},k}$ is determined in a three-step process. Firstly, the mean left- and right-hand pressure-heights are computed, taken as a simple average between the associated layer interfaces in the $i$-th and $(i+1)$-th columns:
\begin{gather}
\label{eqn_cv_gamma_1}
\bar{p}_{i-\frac{1}{2},k} = \frac{1}{2}\left( p_{i,k-\frac{1}{2}} + p_{i,k+\frac{1}{2}} \right),
\qquad
\bar{p}_{i+\frac{1}{2},k} = \frac{1}{2}\left( p_{i+1,k-\frac{1}{2}} + p_{i+1,k+\frac{1}{2}} \right)
\end{gather}
These midpoints define the initial upper and lower surfaces $p_{t},\,p_{b}$ for the control-volume $\Gamma_{i+\frac{1}{2},k}$, such that:
\begin{gather}
p_{t}^{*} = \min\left(\bar{p}_{i-\frac{1}{2},k},\,\bar{p}_{i+\frac{1}{2},k}\right),
\qquad
p_{b}^{*} = \max\left(\bar{p}_{i-\frac{1}{2},k},\,\bar{p}_{i+\frac{1}{2},k}\right)
\end{gather}
A minimum thickness constraint is imposed, ensuring that weakly-sloping layers are inflated to a mean adjacent thickness value:
\begin{gather}
p_{t}^{**} = \min\left(p_{t}^{*},\, \frac{1}{2}\left(p_{t}^{*}+p_{b}^{*}\right) - \frac{\bar{d}}{2} \right),
\qquad
p_{b}^{**} = \max\left(p_{b}^{*},\, \frac{1}{2}\left(p_{t}^{*}+p_{b}^{*}\right) + \frac{\bar{d}}{2} \right),
\\[2ex]
\text{where}\qquad
\bar{d} = \frac{1}{2}\left(\Delta p_{i} + \Delta p_{i+1}\right)
\end{gather}
Finally, these values are limited by the vertical extents of the adjacent fluid columns, ensuring that the control-volumes $\Gamma_{i+\frac{1}{2},k}$ do not protrude either above the fluid surface, or below the bottom boundary:
\begin{gather}
\label{eqn_cv_gamma_2}
p_{t} = \max\left(p_{t}^{**},\, p_{i,\frac{1}{2}},\, p_{i+1,\frac{1}{2}}\right),
\qquad
p_{b} = \min\left(p_{b}^{**},\, p_{i,n_{z}+\frac{1}{2}},\, p_{i+1,n_{z}+\frac{1}{2}}\right)
\end{gather}
Note that such choices are carefully selected to ensure that the control-volumes $\Gamma_{i+\frac{1}{2},k}$ always maintain positive thickness, and that they at least partially overlap with the associated $k$-th layer mass-cells in their adjacent columns, unless an intersection with the fluid surface or bottom bathymetry is encountered.  

\subsection{Evaluation of overlapping integral terms}

\medskip

Recalling the methodology presented in Section~\ref{section_numerical_integration}, the rectilinear pressure gradient force is evaluated as a two-step procedure, firstly seeking to compute the column-wise distributions of geopotential height $\Phi_{i,k}(p)$ through integration of the hydrostatic expression (Eqn.~\ref{eqn_momentum_1}), before evaluating the contact pressure integrals defined in Eqn.~\ref{eqn_pressure_integral_3}. Starting from Eqn.~\ref{eqn_hydro_exact} and using a suitable numerical integration rule, the variation in geopotential height within the $k$-th layer of the $i$-th column is given by:
\begin{gather}
\label{eqn_geo_layer_2}
\Phi_{i,k}(\xi) = \Delta p_{i,k}\,\left(a_{1}\xi + \frac{1}{2}a_{2}\xi^{2} + \dots + \frac{1}{n}a_{n}\xi^{n}\right) + \Phi_{i,k+\frac{1}{2}}
\end{gather}
where $\Phi_{i,k+\frac{1}{2}}$ is the value of the geopotential at the base of the layer and the $a_{l}$'s are the coefficients of the polynomial approximation to $\rho^{-1}$, calculated by sampling the fluid specific-volume over a set of integration points distributed over the layer thickness, as discussed in Section~\ref{section_layerwise_method}.

\begin{figure}[t]

\caption{Detailed representation of the rectilinear pressure gradient force scheme. Geopotential profiles are first computed using numerical integration techniques for each column in the model. Further integration of these profiles along the left- and right-hand edges of the staggered control-volume $\Gamma_{i+\nicefrac{1}{2},k}$ gives the full contact pressure force acting on $\Gamma_{i+\nicefrac{1}{2},k}$. Note that evaluation of the side-integral terms may involve an integration spanning multiple fluid layers -- incorporating the set of grid-cells overlapped by the control-volume $\Gamma_{i+\nicefrac{1}{2},k}$.}

\begin{center}
\includegraphics[width=.60\textwidth]{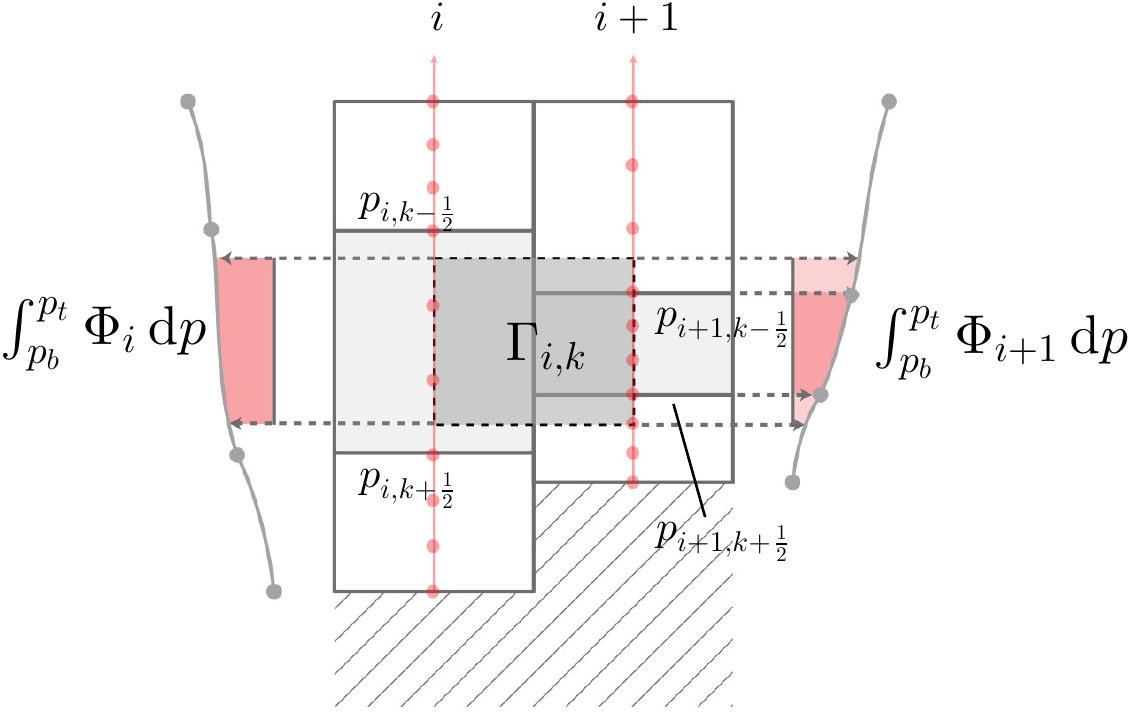}
\end{center}
\end{figure}

Given the variation in $\Phi_{i,k}(p)$ within each column, an evaluation of the contact pressure forces acting along the left- and right-hand edges of the control-volumes $\Gamma_{i+\frac{1}{2},k}$ can be made. Recalling that $\Gamma_{i+\frac{1}{2},k}$ can overlap multiple fluid layers, depending on the relative distribution of layer-thicknesses, the total contact pressure force acting along a given edge is computed as a summation over this set of intersecting layers: 
\begin{gather}
\label{eqn_pgf_interlayer}
\int_{p_{i+\frac{1}{2},k-\frac{1}{2}}}^{p_{i+\frac{1}{2},k+\frac{1}{2}}} \Phi\,\mathrm{d}p = \sum_{q=1}^{|Q|} \Delta p_{i+\frac{1}{2},k}\int_{\xi_{b}}^{\xi_{t}} \Phi\,\mathrm{d}p
\end{gather}
where the set of overlapping layers $Q$ includes any layer $q$ that intersects with the control-volume $\Gamma_{i+\nicefrac{1}{2},k}$ such that $p_{i,q+\nicefrac{1}{2}} \geq p_{i+\nicefrac{1}{2},k-\nicefrac{1}{2}}$ and $p_{i,q-\nicefrac{1}{2}} \leq p_{i+\nicefrac{1}{2},k+\nicefrac{1}{2}}$. Making use of the polynomial form of $\Phi_{i,k}$ given in Eqn.~\ref{eqn_geo_layer_2}, the integrals in Eqn.~\ref{eqn_pgf_interlayer} can be evaluated as follows: 
\begin{gather}
\label{eqn_side_integral_2}
\Delta p_{i+\frac{1}{2},k}\int_{\xi_{b}}^{\xi_{t}} \Phi\,\mathrm{d}p = (\Delta p_{i,k})^{2}\,\left(\frac{1}{2}a_{1}\xi^{2} + \frac{1}{6}a_{2}\xi^{3} + \dots + \frac{1}{n(n+1)}a_{n}\xi^{n+2}\right|_{\xi_{b}}^{\xi_{t}} + \Delta p_{i,k}\,\Phi_{i,k+\frac{1}{2}}
\end{gather}
where $\xi_{t}$ and $\xi_{b}$ are the values of the local coordinate at the endpoints of the intersecting interval. As per Eqn.~\ref{eqn_pgf_interlayer}, the total pressure force acting over the left- and right-hand edges of a control-volume $\Gamma_{i+\nicefrac{1}{2},k}$ is found through a summation of the various integral contributions given by Eqn.~\ref{eqn_side_integral_2}.

\subsection{Summary of rectilinear pressure gradient formulation}

\medskip

The numerical procedure to evaluate the pressure gradient force using the rectilinear finite-volume formulation can be summarised in the following steps:
\begin{enumerate}\itemsep=+4pt
\item Compute the set of piece-wise polynomial reconstructions in the vertical direction for the thermodynamic variables. Specifically, a set of piecewise polynomial interpolants $T_{i,k}(p),\,S_{i,k}(p)$ are computed for each column in the model.

\item Integrate the hydrostatic relationship for the geopotential profiles $\Phi_{i,k}(p)$ associated with each column in the model using Eqn.~\ref{eqn_geo_layer_2}. Integration proceeds layer-by-layer from the base of each column, upwards towards the fluid surface.

\item Evaluate the pressure gradient term for each staggered control-volume $\Gamma_{i+\nicefrac{1}{2},k}$. This is a multi-step process in which: (i) the axis-aligned control-volume $\Gamma_{i+\nicefrac{1}{2},k}$ is formed using Eqn.'s~\ref{eqn_cv_gamma_1}--\ref{eqn_cv_gamma_2}, (ii) the set of intersecting layers $Q_{i+\nicefrac{1}{2},k}$ is computed, by searching for layers in the adjacent columns that overlap with $\Gamma_{i+\nicefrac{1}{2},k}$, and (iii) the contact pressure force acting over the left- and right-hand edges of $\Gamma_{i+\nicefrac{1}{2},k}$ is evaluated using Eqn.~\ref{eqn_pgf_interlayer} and Eqn.~\ref{eqn_side_integral_2}. The subsequent pressure gradient term is taken as the difference in integrated contact pressure over $\Gamma_{i+\nicefrac{1}{2},k}$, as per Eqn.~\ref{eqn_pressure_integral_2}. 
\end{enumerate}
Compared to the layer-wise formulation presented previously, note that the rectilinear scheme is composed entirely of column-centred operations, and does not require horizontal interpolation operations or the computation of geopotential profiles at staggered horizontal points.


\section{Experimental results}

\label{section_layerwise_results}

The performance of the layer-wise and rectilinear finite-volume formulations for evaluation of the horizontal pressure gradient force were assessed using a seres of two-dimensional flow configurations. Specifically, a set of \textit{ocean-at-rest} test-cases were analysed -- measuring the accuracy and consistency of the numerical schemes when subject to increasingly difficult combinations of thermodynamic stratification and layer-wise slope. Specifically, the flows focus on the evolution of a stratified fluid, initialised in equilibrium over a region of rough topography. To provide a stringent test of the numerical formulations, the problem was discretised using a pure terrain-following coordinate -- generating a set of layers of non-uniform thickness, steeply inclined to the horizontal. The fully non-linear TEOS-10 equation-of-state \citep{mcdougall2011getting} was employed in all test cases, as an example of a complex non-linear density function.

Noting that the flow is initialised in equilibrium, the accuracy and consistency of the various pressure gradient formulations can be assessed by measuring the magnitude of \textit{drift} in the flow over time -- analysing the development of both spurious horizontal velocity components and anomalous thermodynamic variations. Schemes that preserve exact hydrostatic consistency are capable of maintaining an unperturbed flow state over time.

\subsection{Initial conditions}

\medskip

A careful initialisation procedure is required to ensure that a correctly equilibrated flow-state is computed with respect to the various non-linearities present in the problem specification. Specifically, interactions between the imposed stratification profiles, layer-wise geometries and equation-of-state definitions are required to be addressed with a high degree of accuracy. Considering a consistent vertical integration of the hydrostatic relationship within each column: 
\begin{gather}
\label{eqn_hydro_pressure_bc}
\partial_{z}(p) = -g\rho\left(T(p),S(p),p\right)
\end{gather}
it is necessary to ensure that: (i) the bottom pressure boundary condition is computed in a sufficiently accurate manner, and (ii) the numerical temperature and salinity degrees-of-freedom are computed for each layer as a consistent integral mean. In this study, Eqn.~\ref{eqn_hydro_pressure_bc} was integrated using a high-order accurate Runge-Kutta type method \citep{shampine1997matlab} over a high-resolution vertical grid. Exact analytic representations of the imposed temperature $T_{0}(p)$ and salinity $S_{0}(p)$ profiles were adopted, allowing an integration of Eqn.~\ref{eqn_hydro_pressure_bc} without additional interpolation considerations. Such a procedure ensures that the discrete bottom pressure boundary condition can be computed to within numerical precision. Additionally, careful initialisation of the grid-cell degrees-of-freedom was employed -- using a high-order accurate numerical integration rule to compute the layer mean quantities:
\begin{gather}
\bar{T}_{i,k} = \frac{1}{\Delta p_{i,k}} \int_{p_{b}}^{p_{t}} T(p)\,\mathrm{d}p,
\qquad
\bar{S}_{i,k} = \frac{1}{\Delta p_{i,k}} \int_{p_{b}}^{p_{t}} S(p)\,\mathrm{d}p
\end{gather}
Again, using the analytic profiles $T_{0}(p)$ and $S_{0}(p)$, such quantities can be computed to within numerical precision by adopting a suitably accurate quadrature rule. Note that such an approach can differ significantly from a simple `midpoint' type approximation to the layer mean values.

\begin{figure}[t]
\centering
\includegraphics[width=.825\textwidth]{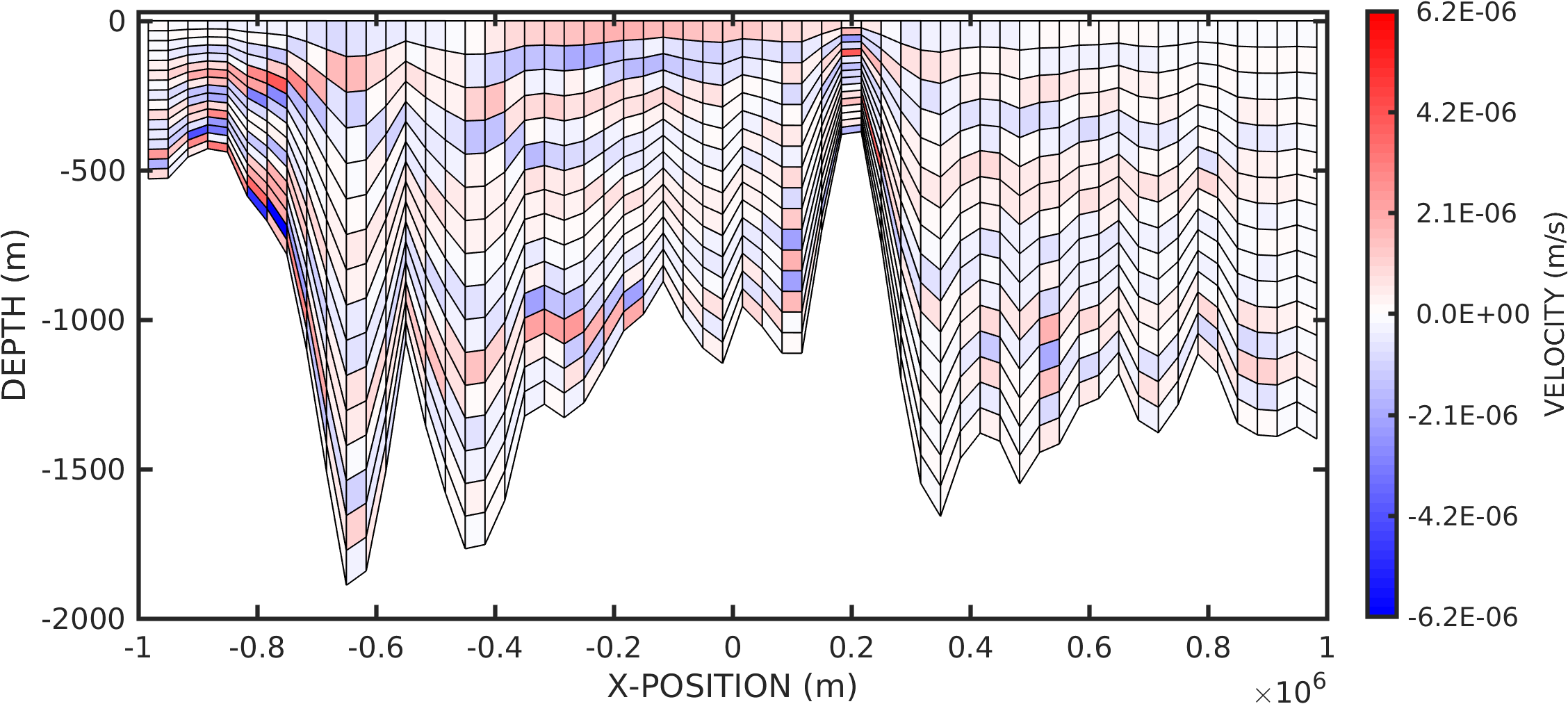}
\\[1ex]
\includegraphics[width=.825\textwidth]{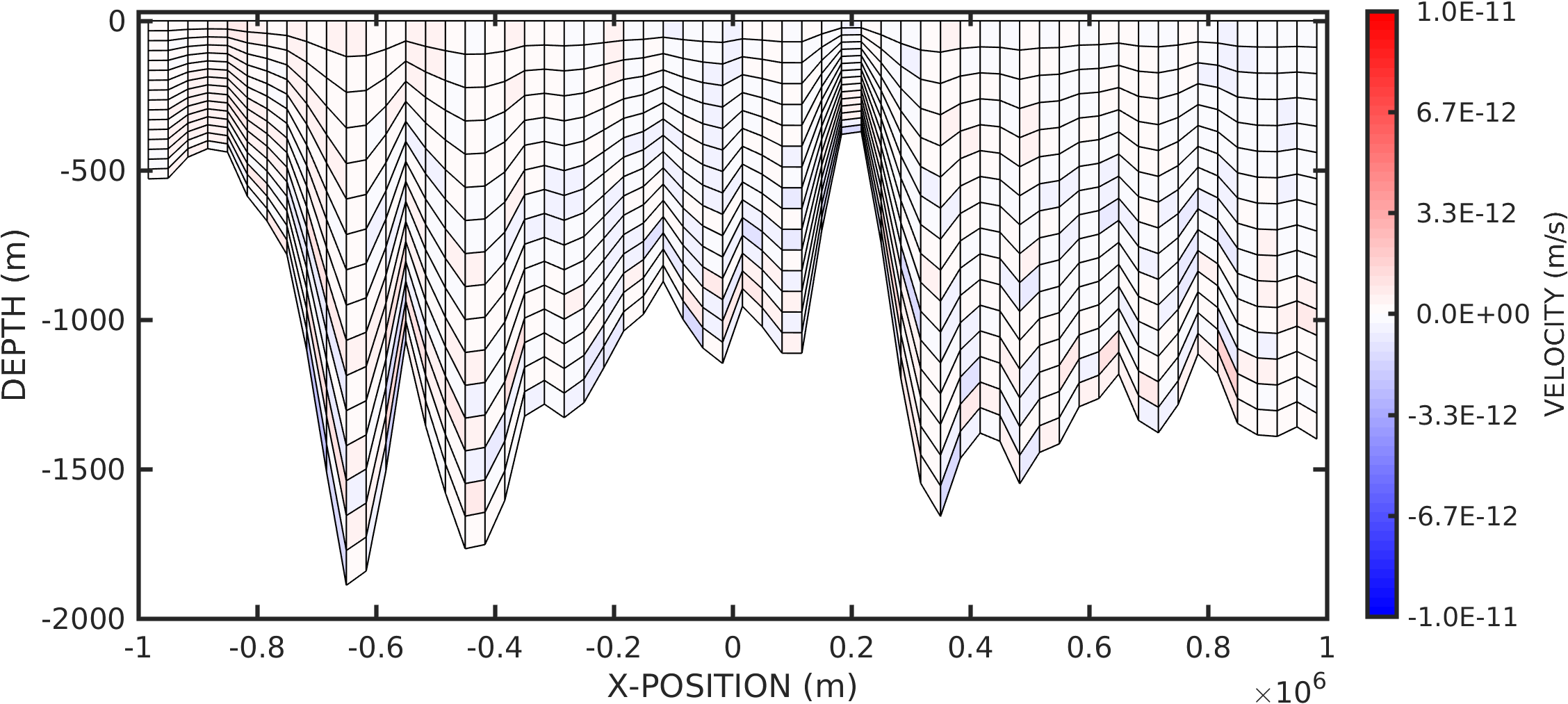}

\caption{Anomalous horizontal velocity magnitude after 90 days of integration using the layer-wise pressure gradient formulation and linear temperature and salinity initial conditions. In the upper panel a $1\times 3$ integration rule is employed, with a $3\times 5$ rule used in the lower panel. Reduced velocity magnitude shows that error approaches zero when a suitably high-order accurate numerical integration procedure is adopted.}

\label{figure_linear_layerwise}

\end{figure}

\subsection{Model setup \& geometry}

\label{section_model_setup}

\medskip

A simple two-dimensional box-model was used for the integration of all flows. The horizontal dimension of the box was set to $2000\,\mathrm{km}$, and was discretised into $60$ uniformly-spaced grid-cells. The vertical axis of the model was configured according to a `pure' sigma-type coordinate, with a stack of $16$ terrain-following layers used in all columns. No warping of coordinate surfaces was incorporated, with the layers within a given column comprising equal thicknesses. The bottom bathymetry was selected to model an environment containing steeply-sloping segments.

The box-model is based on a semi-implicit Arbirary Lagrangian-Eulerian (ALE) type formulation, with the external surface-mode resolved via an implicit operator \citep{marshall1997finite}, and vertical advection achieved via a conservative remapping operation \citep{bleck2002oceanic,white2009high}. Horizontal and vertical advection is accomplished via a high-order accurate essentially monotonic PPM/PQM formulation \citep{engwirda2016weno}. The model time-step was set to $\Delta t=1200$ seconds, with vertical advection activated once every $12$ hours. All flows were integrated over a $90$ day period. No vertical or horizontal mixing schemes were implemented, with explicit dissipation limited to a small horizontal and vertical momentum diffusion operator and frictional bottom boundary condition.

\subsection{Linear stratification}

\begin{figure}[t]
\centering
\includegraphics[width=.825\textwidth]{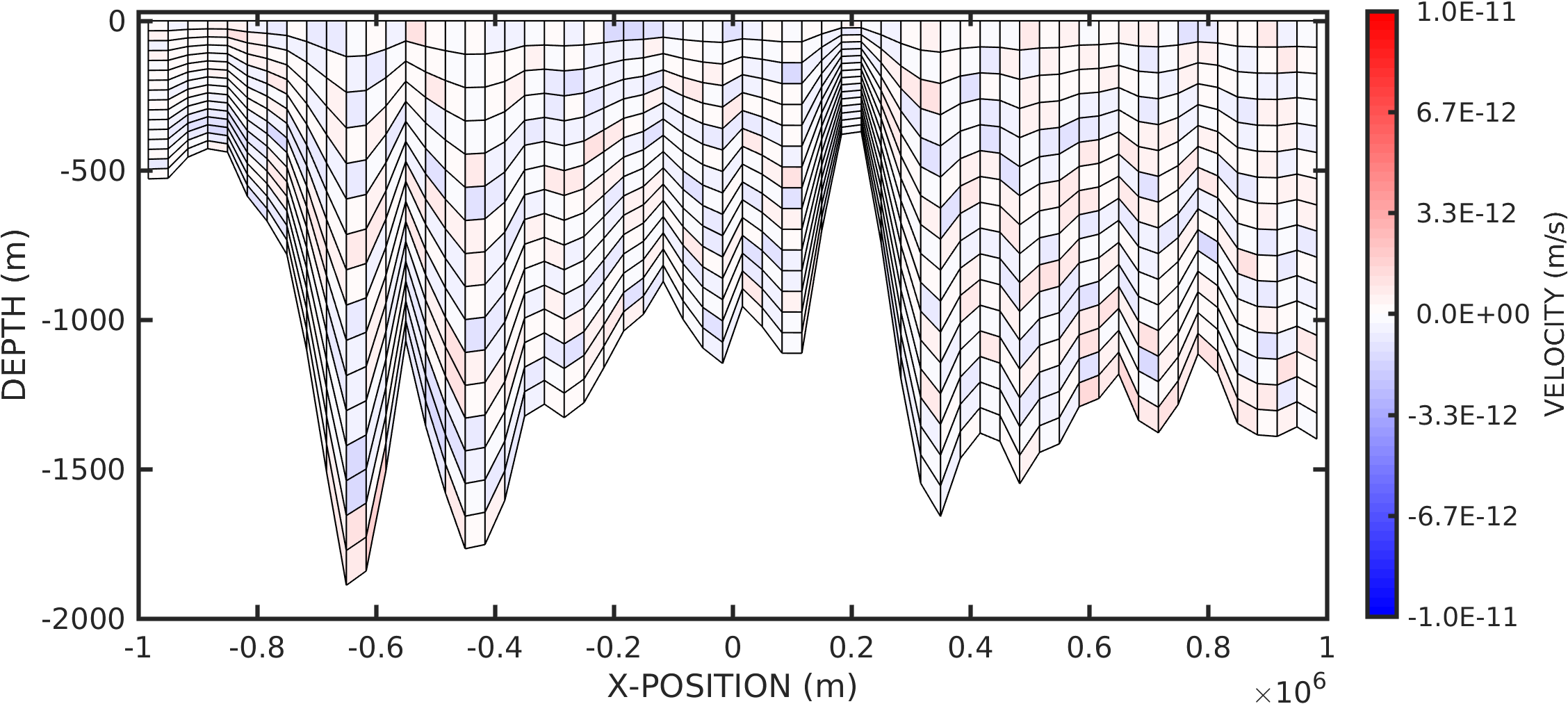}

\caption{Anomalous horizontal velocity magnitude after 90 days of integration using the rectilinear pressure gradient formulation and linear temperature and salinity initial conditions. Consistent with the layer-wise results, use of high-order accurate, 5-point integration rule preserves hydrostatic consistency to machine precision.}

\label{figure_linear_rectilinear}

\end{figure}

\medskip

In the first test problem, the fluid was equilibrated using a set of linear temperature and salinity initial conditions:
\begin{gather}
\label{eqn_linear_profile}
T_{0}(p) = T_{s} - \left(\frac{\Delta T}{\Delta p}\right) p,
\qquad
S_{0}(p) = S_{s} + \left(\frac{\Delta S}{\Delta p}\right) p
\end{gather}
where $\Delta p = 2 \times 10^{7}\,\mathrm{Pa}$, $T_{s}=20\, ^\circ\mathrm{C}$, $\Delta T = 20\, ^\circ\mathrm{C}$, $S_{s}=10\,\nicefrac{\mathrm{g}}{\mathrm{kg}}$ and $\Delta S=25\,\nicefrac{\mathrm{g}}{\mathrm{kg}}$. Such profiles give temperatures and salinities of $T=20\, ^\circ\mathrm{C}$, $S=10\,\nicefrac{\mathrm{g}}{\mathrm{kg}}$ at the fluid surface, and $T=0\, ^\circ\mathrm{C}$, $S=35\,\nicefrac{\mathrm{g}}{\mathrm{kg}}$ at the lowest point on the bottom boundary.

Firstly, the convergence of both the layer-wise and rectilinear formulations was assessed, by varying the order of the numerical integration rules used to compute the pressure gradient force. In Figure~\ref{figure_linear_layerwise}, the horizontal velocity field after 90 days of integration using the layer-wise formulation is shown. In the top panel, results using a `low-order' pressure gradient scheme are illustrated, in which a so-called $1 \times 3$ integration rule is used, employing one integration point in the vertical and three in the horizontal. An analysis of the velocity field shows a relatively small spurious horizontal flow, with a maximum magnitude of approximately $6 \times 10^{-6}\, \nicefrac{\mathrm{m}}{\mathrm{s}}$. In the bottom panel, results using a `higher-order' pressure gradient scheme are shown, in which a $3 \times 5$ integration rule is used, employing three integration points in the vertical and five in the horizontal. The associated spurious velocity field shows a maximum error of less than $1 \times 10^{-11}\, \nicefrac{\mathrm{m}}{\mathrm{s}}$ in this case, demonstrating that the layer-wise pressure gradient formulation -- when based on sufficiently high-order accurate numerical integration rules -- leads to an essentially \textit{error-free} discretisation for this test-case, with hydrostatic equilibrium maintained to machine precision. A similar experiment was conducted for the rectilinear formulation, leading to comparable conclusions. Specifically, it was found that use of a sufficiently high-order accurate, 5-point integration rule led to essentially error-free behaviour, with maximum spurious velocity currents of less than $1 \times 10^{-11}\, \nicefrac{\mathrm{m}}{\mathrm{s}}$ reported after 90 days of integration. See Figure~\ref{figure_linear_rectilinear} for additional details and contours. 

The ability to represent ocean states incorporating linear stratification profiles and arbitrary non-linear equation-of-state definitions represents an improvement on the original semi-analytic scheme of \citet{adcroft2008finite} which was limited to piecewise constant thermodynamic profiles and a simplified fluid density function \citep{wright1997equation}. Though an imposed linear stratification profile may initially seem innocuous, it should be noted that significant non-linearities -- due to both thermodynamic and pressure-compressibility effects are encountered even in this simple case, when the complexities of a fully non-linear equation-of-state are considered \citep{mcdougall2011getting}.

\subsection{Quadratic stratification}

\begin{figure}[t]
\centering
\includegraphics[width=.825\textwidth]{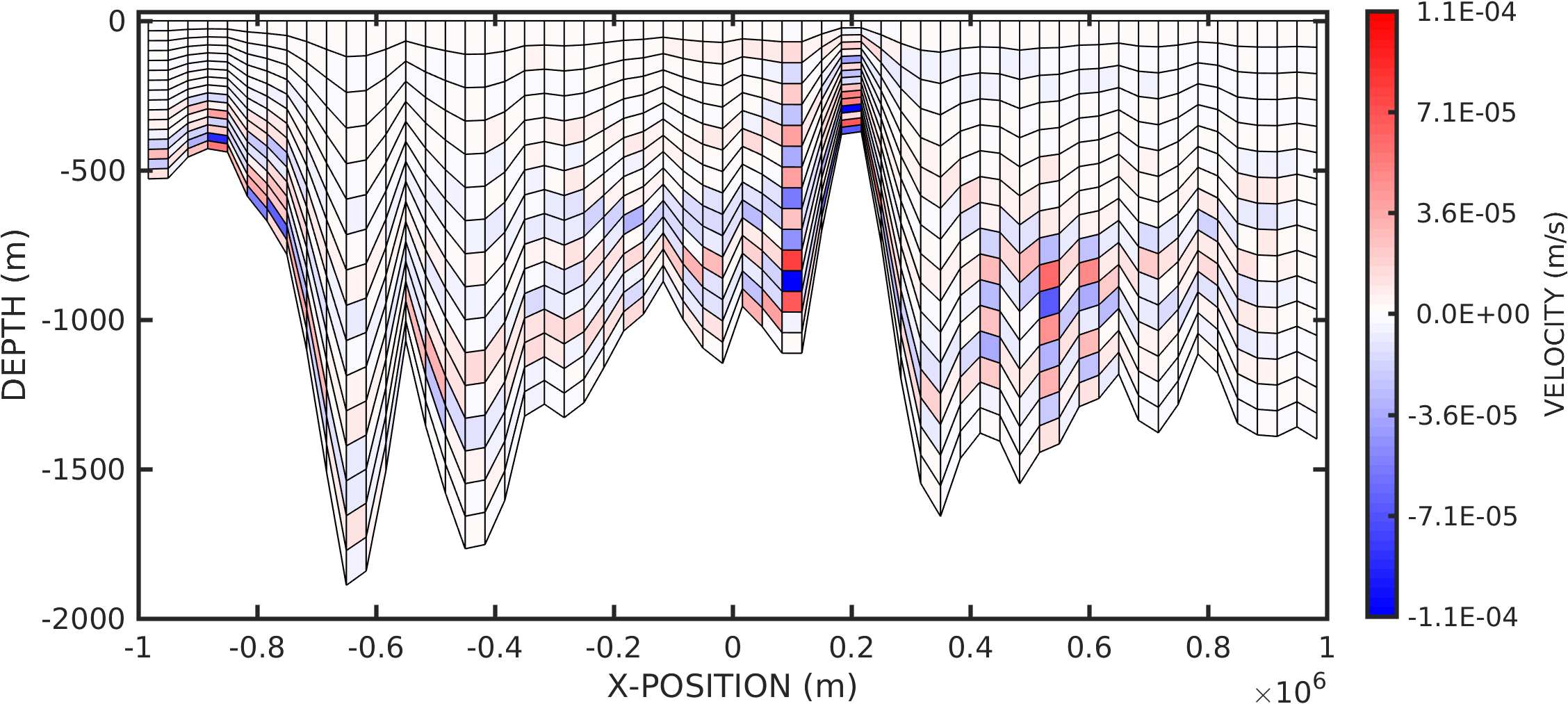}
\\[1ex]
\includegraphics[width=.825\textwidth]{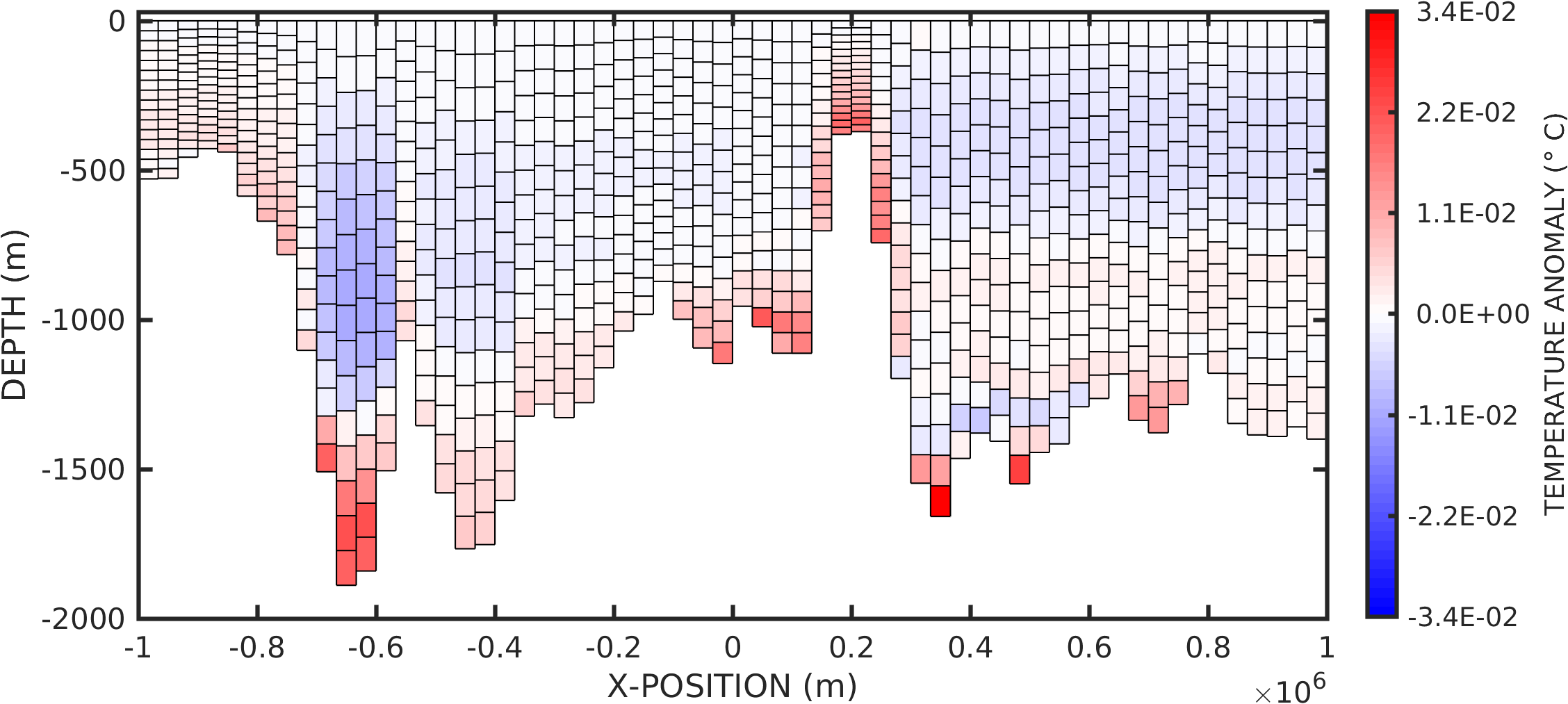}

\caption{Anomalous horizontal velocity magnitude and temperature anomaly after 90 days of integration using the layerwise pressure gradient formulation and quadratic temperature initial conditions. Spurious velocity currents and thermal drift patterns are clearly evident.}

\label{figure_quadratic_layerwise}

\end{figure}

\begin{figure}[t]
\centering
\includegraphics[width=.825\textwidth]{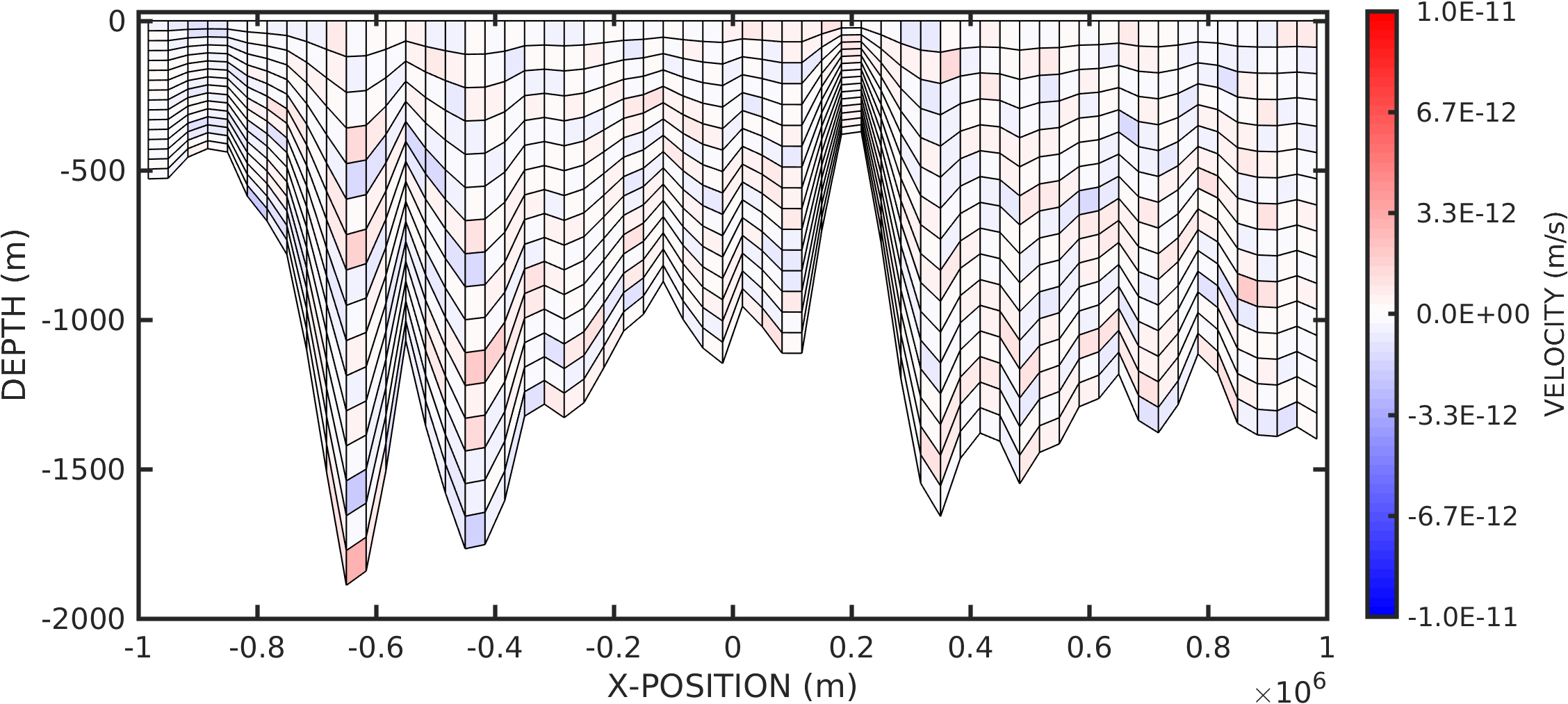}
\\[1ex]
\includegraphics[width=.825\textwidth]{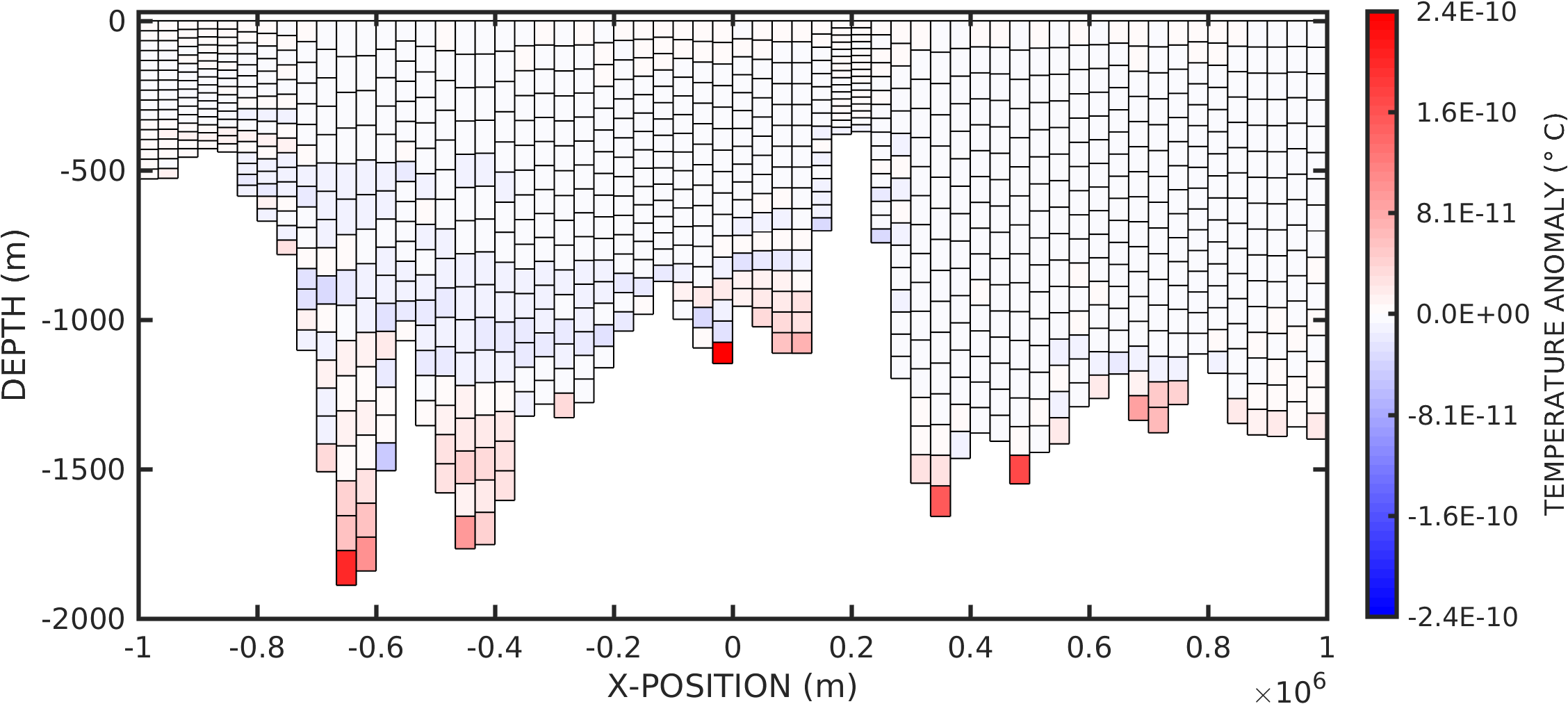}

\caption{Anomalous horizontal velocity magnitude and temperature anomaly after 90 days of integration using the rectilinear pressure gradient formulation and quadratic temperature initial conditions. The flow is preserved in an essentially error-free fashion.}

\label{figure_quadratic_rectilinear}

\end{figure}

\begin{figure}[t]
\centering
\includegraphics[width=.825\textwidth]{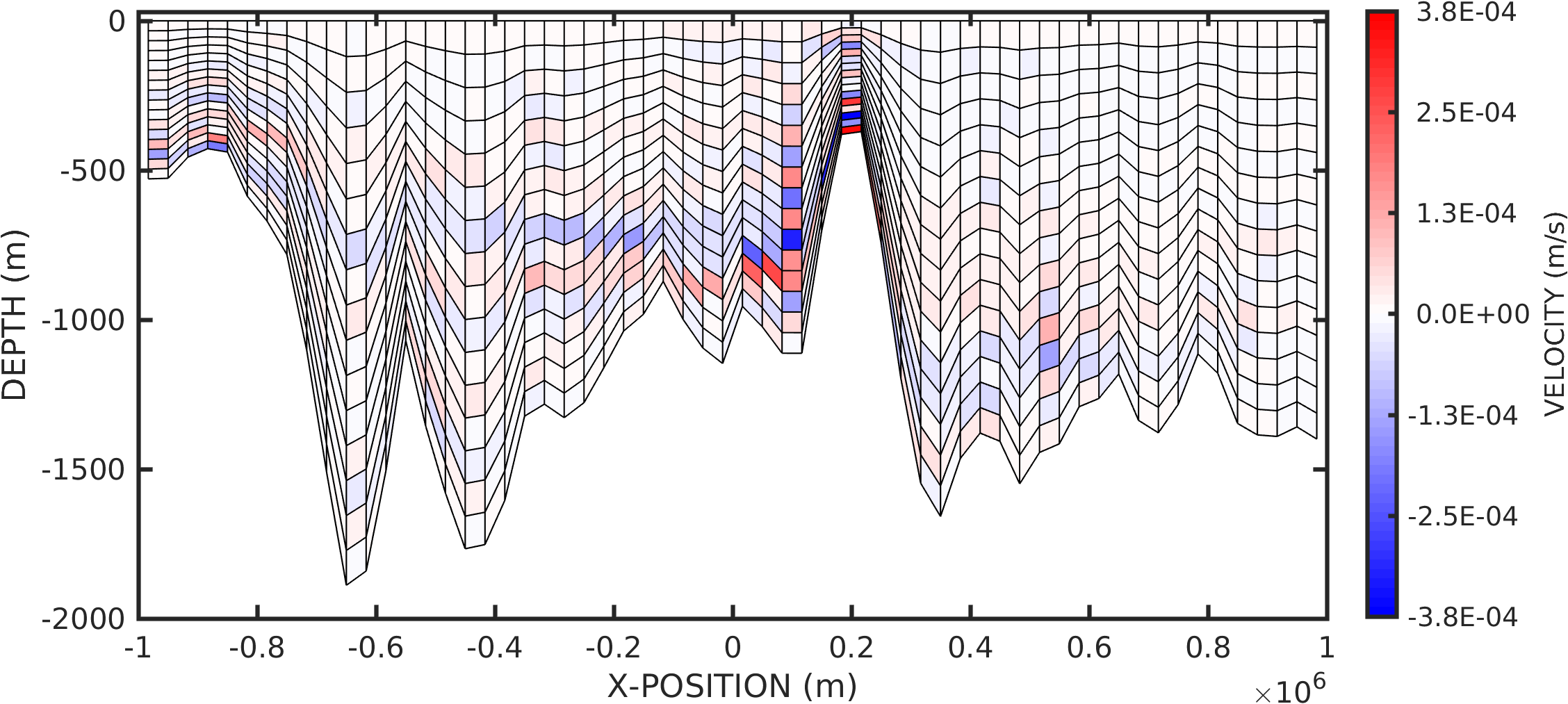}
\\[1ex]
\includegraphics[width=.825\textwidth]{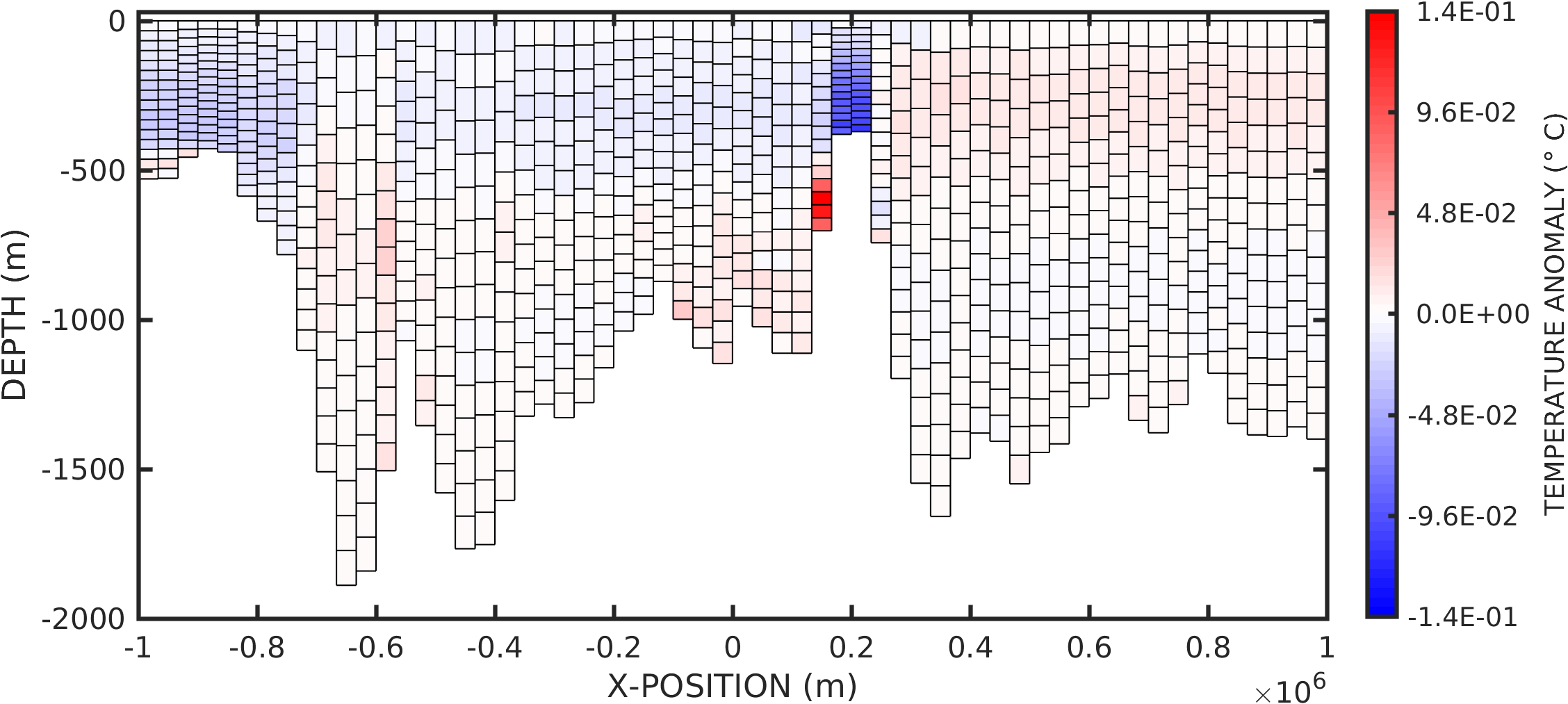}

\caption{Anomalous horizontal velocity magnitude and temperature anomaly after 90 days of integration using the layerwise pressure gradient formulation and exponential temperature initial conditions. Spurious velocity currents and thermal drift patterns are clearly evident.}

\label{figure_exponential_layerwise}

\end{figure}

\begin{figure}[t]
\centering
\includegraphics[width=.825\textwidth]{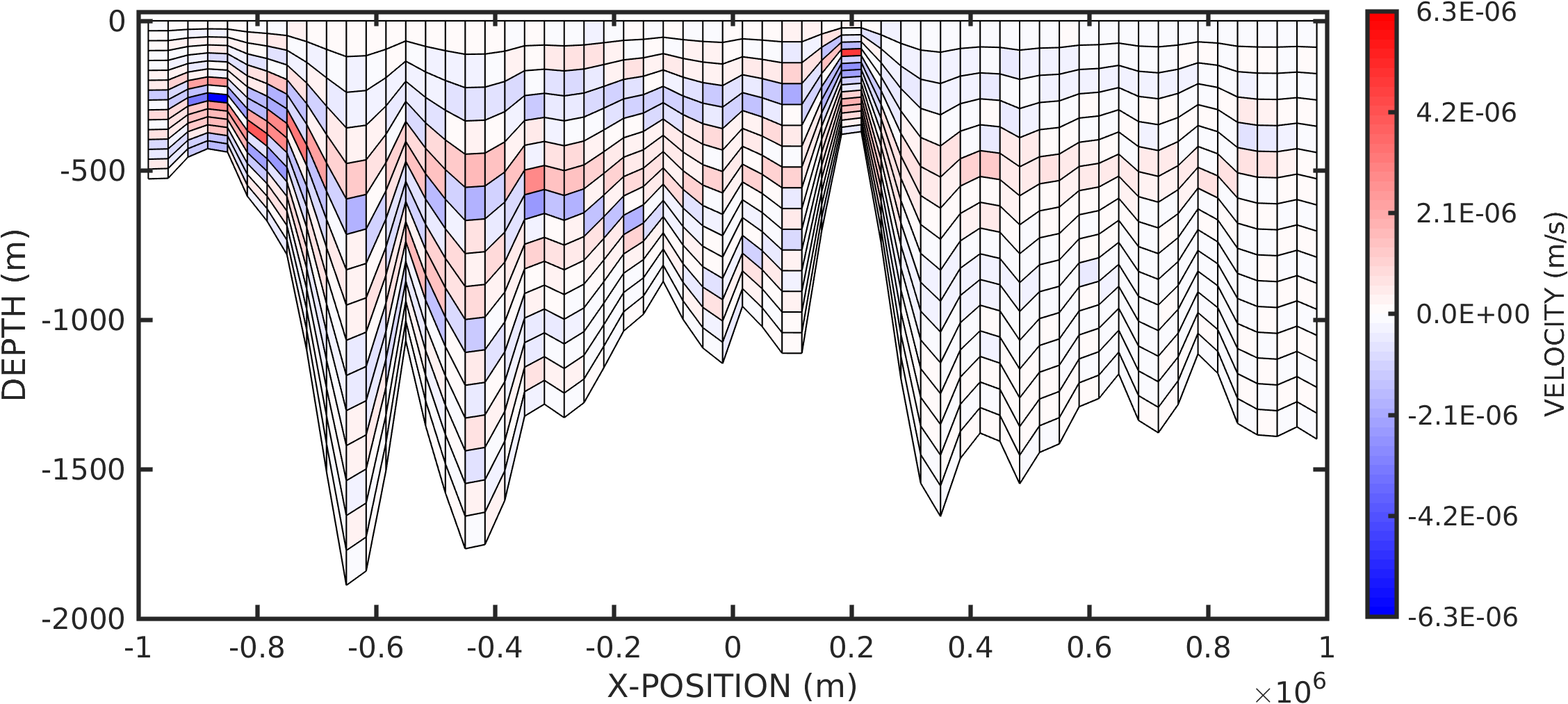}
\\[1ex]
\includegraphics[width=.825\textwidth]{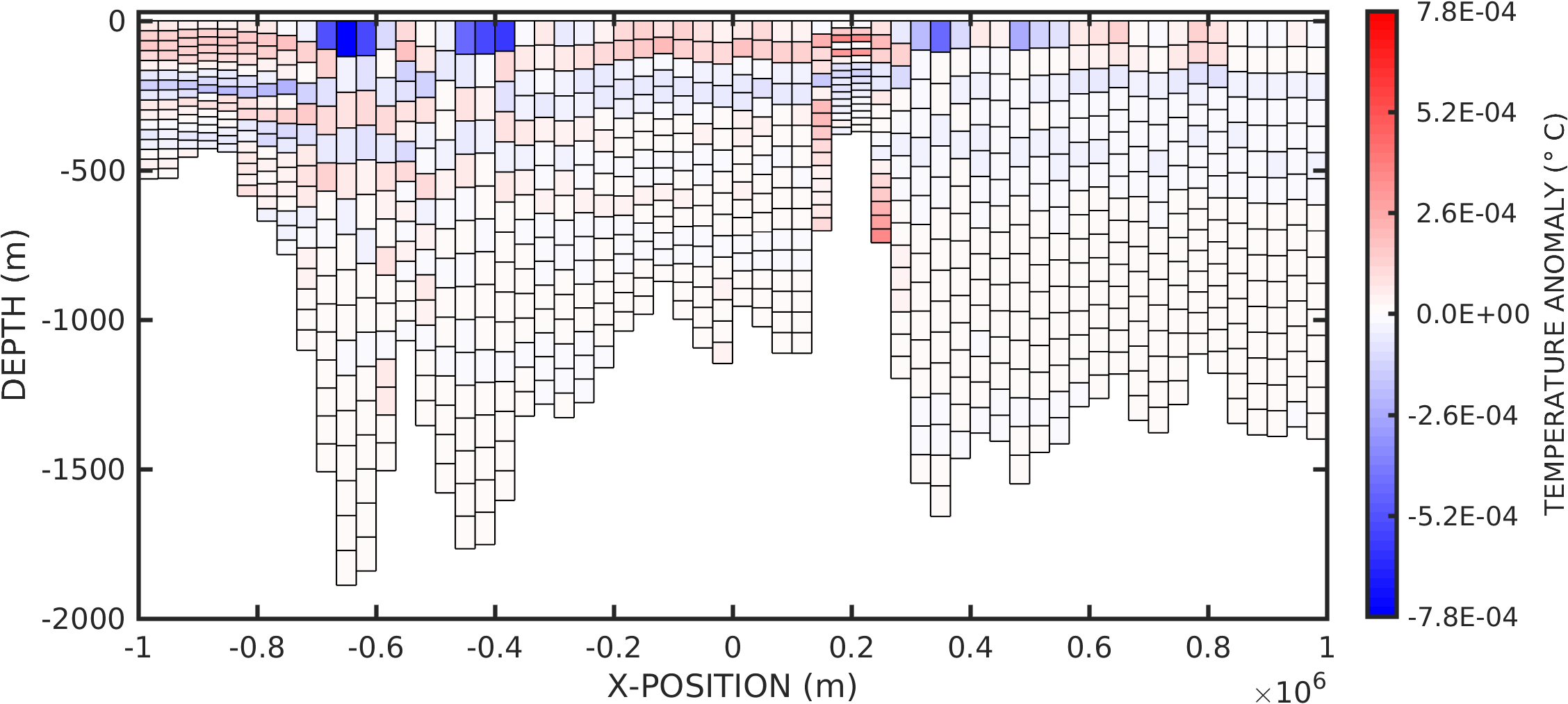}

\caption{Anomalous horizontal velocity magnitude and temperature anomaly after 90 days of integration using the rectilinear pressure gradient formulation and exponential temperature initial conditions. Spurious velocity currents and thermal drift patterns are clearly evident.}

\label{figure_exponential_rectilinear}

\end{figure}

\medskip

In the second test problem, the influence of non-linear thermodynamic stratification was assessed, with the fluid equilibrated using a set of quadratic temperature and linear salinity initial conditions:
\begin{gather}
\label{eqn_quadratic_profile}
T_{0}(p) = T_{s} - \left(\frac{\Delta T}{\Delta p}\right) p - \left(\frac{\Delta T'}{\Delta p}\right)^{2} p^{2},
\qquad
S_{0}(p) = S_{s} + \left(\frac{\Delta S}{\Delta p}\right) p
\end{gather}
where, in addition to those constants defined previously in Eqn.~\ref{eqn_linear_profile}, $\Delta T'=\sqrt{10}$ and $T_{s}=30\, ^\circ\mathrm{C}$. Note that the imposed temperature and salinity profiles $T_{0}(p)$ and $S_{0}(p)$ can be exactly reconstructed using the high-order PPM/PQM interpolants employed in this study. 

Following the results of the linear-profile test-case, both the layer-wise and rectilinear pressure gradient formulations were run using the high-order accurate $3 \times 5$ and $5$-point integration rules, respectively. In Figures~\ref{figure_quadratic_layerwise} and \ref{figure_quadratic_rectilinear}, contours of the horizontal velocity field and anomalous temperature distribution are shown after 90 days of integration. Focusing firstly on Figure~\ref{figure_quadratic_layerwise}, it can be seen that the layer-wise formulation fails to maintain exact consistency in this case, with a small spurious velocity component seen to drive an anomalous thermal drift. Specifically, spurious currents on the order $1 \times 10^{-4}\, \nicefrac{\mathrm{m}}{\mathrm{s}}$ are generated, resulting in temperature drifts of approximately $3 \times 10^{-2}\, ^\circ\mathrm{C}$. Errors are seen to be clustered adjacent to layers of significant slope. The absolute magnitude of these errors was not observed to grow with time. 

The genesis of these errors appears to be tied to a single operation embedded within the layer-wise formulation -- specifically, the horizontal interpolation of temperature and salinity profiles  to integration points interior to the control-volumes $\Omega_{i+\nicefrac{1}{2},k}$. Such calculations are necessary when computing intermediate profiles of geopotential $\Phi_{l,k}$, and, subsequently, the contact pressure forces acting along the sloping upper and lower edges of the grid-cell. For control-volumes of significant geometrical slope, the difference between the imposed quadratic temperature profile and a layer-wise linear approximation can become non-negligible -- leading to an erroneous approximation of the pressure forces acting on the sloping interfaces. It is emblematic of the sensitivity of the pressure gradient term itself that such small discrepancies can lead to relative large errors.

In Figure~\ref{figure_quadratic_rectilinear}, results for the rectilinear formulation are presented, and show much improved performance. Specifically, it is seen that essentially error-free behaviour is achieved, with maximum velocity magnitudes of less than $1 \times 10^{-11}\, \nicefrac{\mathrm{m}}{\mathrm{s}}$, reported, inducing negligible thermal drifts of approximately $2 \times 10^{-10}\, ^\circ\mathrm{C}$. These results confirm that, due to the absence of layer-wise interpolation operations, the rectilinear formulation is able to maintain near-perfect hydrostatic consistency in the presence of non-linear stratification profiles, steeply-sloping layer geometries and a complex non-linear equation-of-state definition. Note that in addition to a highly accurate integration of the contact pressure integrals, such behaviour relies on an exact vertical reconstruction of column-wise temperature and salinity profiles. This is achieved in the case of polynomial stratification profiles by making use of high-order accurate PPM/PQM type interpolation schemes.

\subsection{Exponential stratification}

\medskip

In the third test problem, the influence of inexact vertical reconstruction was examined, with the fluid equilibrated using a set of exponential temperature and linear salinity initial conditions:
\begin{gather}
\label{eqn_exponential_profile}
T_{0}(p) = T_{s}\, \mathrm{e}^{-\left(\frac{p-\alpha}{\beta}\right)},
\qquad
S_{0}(p) = S_{s} + \left(\frac{\Delta S}{\Delta p}\right) p
\end{gather}
where, in addition to those constants defined previously in Eqn.~\ref{eqn_linear_profile}, $\alpha = 1 \times 10^{5}\,\mathrm{Pa}$ and $\beta = 5 \times 10^{6}\,\mathrm{Pa}$. Note that, in contrast to the previous test-cases, the imposed temperature profile $T_{0}(p)$ cannot be exactly reconstructed using the polynomial-type PPM/PQM interpolants employed in this study. 

Consistent with previous test-cases, the layer-wise and rectilinear pressure gradient formulations were run using the $3 \times 5$ and $5$-point integration rules, respectively. In Figures~\ref{figure_exponential_layerwise} and \ref{figure_exponential_rectilinear}, contours of the horizontal velocity field and anomalous temperature distribution are shown after 90 days of integration. In this case, both pressure gradient formulations are seen to exhibit some level of spurious movement, though the errors associated with the layer-wise method are almost two orders of magnitude larger than those associated with the rectilinear scheme. Specifically, the layer-wise method induces spurious currents on the order of $4 \times 10^{-4}\, \nicefrac{\mathrm{m}}{\mathrm{s}}$, leading to a maximum thermal drift of approximately $0.15\, ^\circ\mathrm{C}$. For the rectilinear formulation, a maximum spurious velocity component of $6 \times 10^{-6}\, \nicefrac{\mathrm{m}}{\mathrm{s}}$ is reported, associated with a thermal drift of approximately $8 \times 10^{-4}\, ^\circ\mathrm{C}$. In both cases, it was observed that the absolute magnitude of these errors did not grow with time.

The source of the pressure gradient errors in this test-case are two-fold. Firstly, consistent with observations made in the previous test problem, errors in the layer-wise formulation can be attributed primarily to the action of the horizontal interpolation scheme used to evaluate temperature and salinity at interior integration points. This assumption is reinforced by noting that the magnitude of the spurious velocity components in both the `exponential' and `quadratic' test-cases are of the same order when the layer-wise scheme is used. Furthermore, errors are seen to be clustered in areas of significant layer slope.

Additionally, there also exist a set of lower-order errors due to an inexact vertical reconstruction of the imposed exponential temperature profiles. An analysis of Figure~\ref{figure_exponential_rectilinear}, shows that errors associated with the rectilinear scheme are concentrated near the surface layers, primarily adjacent to grid-cells of larger thickness. Noting, firstly, that the gradient of the imposed exponential profile is largest at the surface, and secondly, that lower accuracy one-sided polynomial approximations are employed by the PPM/PQM interpolation schemes in grid-cells adjacent to boundaries, it is argued that such errors are a by-product of the vertical interpolation scheme. It was found that by switching from the 3rd-order accurate PPM interpolant (results shown in Figure~\ref{figure_exponential_rectilinear}) to the 5th-order accurate PQM scheme (results not shown), the magnitude of the spurious horizontal velocity was reduced by more than an order of magnitude. Such results highlight the benefits of employing sufficiently high-order accurate reconstruction techniques. 


\section{Discussion \& Conclusions}

\label{section_conclusion}

A pair of finite-volume formulations for evaluation of the horizontal pressure gradient force in layered ocean models have been presented. Through the use of high-order accurate numerical quadrature and polynomial reconstruction techniques, both methods have been designed to maintain hydrostatic and thermobaric equilibrium in the presence of strongly-sloping layer-wise geometries, non-linear equation-of-state descriptions and non-uniform vertical stratification profiles. The use of high-order accurate numerical integration procedures can be seen as a generalisation of previous finite-volume type approaches \citep{adcroft2008finite}. The two formulations differ primarily in their choice of staggered momentum control-volumes, with the layer-wise method based on a conforming, piecewise linear interpolation of adjacent column-wise pressure-levels, while the rectilinear method employs an axis-aligned geometry that may overlap multiple adjacent fluid layers.  

The performance of the new schemes was assessed using a set of two-dimensional benchmark problems, designed to measure the dynamical `drift' away from a non-linear equilibrium state over time. Overall, both methods were shown to perform well -- able to achieve exact consistency in the presence of steeply-sloping terrain-following layers, a complex, non-linear equation-of-state definition, and linear vertical stratification profiles. In the presence of more complex thermodynamic configurations, the rectilinear method was shown to outperform the layer-wise formulation. Specifically, it was found that the horizontal interpolation operator embedded within the layer-wise formulation can lead to erroneous pressure gradient force evaluations when the imposed stratification profiles are non-linear and the layers steeply-sloped. While the construction of higher-order accurate interpolation procedures seems an obvious improvement, the development of such techniques is not necessarily trivial, due to the difference in orientation between the curvilinear layers and true horizontal axis.

The performance of the rectilinear formulation appears to be particularly promising, with this method leading to either exact, or highly accurate pressure gradient force evaluations for all test-cases analysed. Further study is required to assess the behaviour of this scheme in a fully dynamic context, and in a coupled, three-dimensional global ocean environment.

\section*{Acknowledgements}

This work was conducted at the NASA Goddard Institute for Space Studies (NASA-GISS), and was supported by a NASA-GISS / MIT cooperative research agreement.

\appendix
\section{Numerical Integration Coefficients}
\label{appendix_int}

A set of high-order accurate numerical integration rules for evaluation of the hydrostatic and contact pressure integrals can be derived using standard numerical quadrature techniques. Specifically, noting that integrals involving both the geopotential:
\begin{gather}
\label{eqn_hydro_integral_appendix}
\begin{split}
\Phi_{i,k}(p)-\Phi_{i,k+\frac{1}{2}} = \int_{p_{i,k+\frac{1}{2}}}^{p} \rho^{-1}\,\mathrm{d}p 
&\simeq \Delta p\int_{0}^{\xi} a_{1} + a_{2}\xi + \dots + a_{n}\xi^{n-1}\,\mathrm{d}\xi
\\[1ex]
&\simeq \Delta p\,\left(a_{1}\xi + \frac{1}{2}a_{2}\xi^{2} + \dots + \frac{1}{n}a_{n}\xi^{n}\right)
\end{split}
\end{gather}
and contact pressure force:
\begin{gather}
\label{eqn_side_integral_appendix}
\int_{p_{i,k-\frac{1}{2}}}^{p_{i,k+\frac{1}{2}}}\Phi\,\mathrm{d}p = (\Delta p)^{2}\,\left(\frac{1}{2}a_{1} + \frac{1}{6}a_{2} + \dots + \frac{1}{n(n+1)}a_{n}\right) + \Delta p\,\Phi_{i,k+\frac{1}{2}}
\end{gather}
can be evaluated to arbitrarily high orders of accuracy by finding a suitable polynomial expansion, the task is to compute the expansion coefficients $a_{1},a_{2},\dots,a_{n}$ for a given equation-of-state definition and thermodynamic profile. This curve fitting procedure can be accomplished by sampling the integrand (the fluid specific-volume $\nicefrac{1}{\rho}$) at a set of \textit{quadrature-points} distributed over the integration segment. Adopting a standard $n$-term polynomial expansion $f(\xi)$:
\begin{gather}
f(\xi) = \mathbf{b}\, \mathbf{\hat{a}}^{\mathrm{T}}, 
\quad\text{where}\quad 
\mathbf{b} = \left[1,\, \xi,\, \xi^{2},\, \dots,\, \xi^{n-1} \right]
\quad\text{and}\quad
\mathbf{\hat{a}} = \left[a_{1},\, a_{2},\, a_{3},\, \dots,\, a_{n} \right],
\end{gather}
the coefficients $\mathbf{\hat{a}}$ can be evaluated by solving the system of linear equations defined by the interpolation problem:
\begin{gather}
\label{eqn_interp_matrix_appendix}
\left[
\begin{matrix}
1,\, \xi_{1},\, \dots,\, \xi_{1}^{n-1}\\[4pt]
1,\, \xi_{2},\, \dots,\, \xi_{2}^{n-1}\\[4pt]
\vdots\\[4pt]
1,\, \xi_{n},\, \dots,\, \xi_{n}^{n-1}\\[4pt]
\end{matrix}
\right]\,
\left[
\begin{matrix}
a_{1}\\[4pt]
a_{2}\\[4pt]
\vdots\\[4pt]
a_{n}\\[4pt]
\end{matrix}
\right] 
=
\left[
\begin{matrix}
\rho^{-1}\Big(T(\xi_{1}),S(\xi_{1}),p(\xi_{1})\Big)\\[4pt]
\rho^{-1}\Big(T(\xi_{2}),S(\xi_{2}),p(\xi_{2})\Big)\\[4pt]
\vdots\\[4pt]
\rho^{-1}\Big(T(\xi_{n}),S(\xi_{n}),p(\xi_{n})\Big)\\[4pt]
\end{matrix}
\right]
\end{gather}
such that $\mathbf{\hat{a}} = \mathbf{R}^{-1}\, \mathbf{v}$, where $\mathbf{R}^{-1}$ is the matrix of quadrature coefficients, pre-computed for each integration rule as the inverse of the matrix operator in Eqn~\ref{eqn_interp_matrix_appendix}, and $\mathbf{v}$ is the vector of specific-volume evaluations, calculated once for each segment to be integrated. Note that computation of $\mathbf{v}$ requires an evaluation of the quantities $T(\xi_{l})$ and $S(\xi_{l})$ at the sampling points $\xi_{l}$. In this work, such terms are evaluated using an \textit{essentially-monotonic} variant of the 3rd- and 5th-order accurate PPM/PQM interpolants \citep{engwirda2016weno}.  

Optimal sets of sampling points $\xi_{1},\, \xi_{2},\, \dots,\, \xi_{n}$ can be obtained from standard quadrature techniques. For example, 4-point Gauss-Legendre and Lobatto type integration rules can be obtained via:
\begin{gather}
\xi_{\mathrm{G}}^{\,4} = 
\left[
\begin{matrix}
\frac{1}{2} - \frac{1}{2}\sqrt{\frac{1}{7}\alpha_{1}}\\[4pt]
\frac{1}{2} - \frac{1}{2}\sqrt{\frac{1}{7}\alpha_{2}}\\[4pt]
\frac{1}{2} + \frac{1}{2}\sqrt{\frac{1}{7}\alpha_{2}}\\[4pt]
\frac{1}{2} + \frac{1}{2}\sqrt{\frac{1}{7}\alpha_{1}}\\[4pt]
\end{matrix}
\right]
\quad\text{and}\quad
\xi_{\mathrm{L}}^{4} = 
\left[
\begin{matrix}
0\\[4pt]
\frac{1}{2} - \frac{1}{2}\sqrt{\frac{1}{2}}\\[4pt]
\frac{1}{2} + \frac{1}{2}\sqrt{\frac{1}{2}}\\[4pt]
1\\[2pt]
\end{matrix}
\right],
\quad\text{where}\quad
\begin{array}{l}
\alpha_{1} = 3 + 2\sqrt{\tfrac{6}{5}},\\[12pt]
\alpha_{2} = 3 - 2\sqrt{\tfrac{6}{5}}.
\end{array}
\end{gather}
Here $\xi_{\mathrm{G}}^{4}$ and $\xi_{\mathrm{L}}^{4}$ have been mapped onto the unit segment $\xi\in [0,1]$. See, for instance, \citet{golub1969calculation} and \citet{abramowitz1964handbook} for details of additional integration rules.

It is important to note that by computing the full matrix of polynomial expansion coefficients explicitly in this work (Eqn.~\ref{eqn_interp_matrix_appendix}), efficient schemes for the evaluation of the nested geopotential and contact pressure integrals can be formulated using only a single set of equation-of-state evaluations per segment, even in the case of partial or overlapping segments as per the rectilinear formulation. The techniques presented here are otherwise equivalent to standard numerical quadrature rules. 

\section*{References}

\bibliographystyle{elsarticle-harv}
\bibliography{references}

\end{document}